\documentclass[11pt,singlespace]{article}
\usepackage{graphicx}
\usepackage{subfigure}
\usepackage{amsmath}
\newcommand\nomenclature[2]{{#1} & -- & {#2}}%
\DeclareMathAlphabet{\mathsfsl}{OT1}{cmss}{m}{sl}

\DeclareMathOperator{\trace}{tr}

\newcommand{\fepx}{{\bfseries{\slshape{FEpX}}}}
\newcommand{\odfpf}{{\bfseries{\slshape{OdfPf}}}}
\newcommand{\neper}{{\bfseries{\slshape{Neper}}}}
\newcommand{\paraview}{{\bfseries{\slshape{Paraview}}}}
\newcommand{\visit}{{\bfseries{\slshape{VisIt}}}}

\newcommand{\matlab}{{\bfseries{\slshape{MATLAB}}}}

\newcommand{\vctr}[1]{\boldsymbol{#1}}

\newcommand{\tnsr}[1]{\mathsfsl{#1 }}

\newcommand{\gtnsr}[1]{\mathsfsl{#1}}

\newcommand{\rodvec}{\vctr{r}}
\newcommand{\rodvel}{\vctr{v}}
\newcommand{\quaternion}{\vctr{q}}

\newcommand{\cofq}{\boldsymbol{\cal{C}}(\quaternion)}

\newcommand{\mofq}{\boldsymbol{\cal{M}}(\vctr{q},\gammadot)}

\newcommand{\gradop}{{\rm grad }}
\newcommand{\divop}{{\rm div}}

\newcommand{\detop}{{\rm det}}
\newcommand{\vectop}{{\rm vect}}
\newcommand{\transp}{{\rm T}}
\newcommand{\invrs}{{ -\hspace{-.1em}1}}
\newcommand{\deeteei }{ \frac{1}{\Delta t}  }   
\newcommand{\betaoverkappa}{ \frac{\beta}{\kappa \Delta t}  }

\newcommand{\cauchy}{{\boldsymbol{\sigma}}}
\newcommand{\dcauchy}{{\boldsymbol{\sigma}}^\prime}

\newcommand{\kirch}{{\boldsymbol{\tau}}}
\newcommand{\dkirch}{{\boldsymbol{\tau}}^\prime}
\newcommand{\xdefgrad}{\tnsr{f}}
\newcommand{\xdefgraddot}{\dot\xdefgrad}
\newcommand{\xdefgradi}{\tnsr{f}^\invrs}
\newcommand{\pxdefgrad}{{\tnsr{f}}^p}
\newcommand{\latstretch}{\tnsr{v}^e}
\newcommand{\rstar}{\tnsr{r}^\star}
\newcommand{\rstardot}{{\dot{\tnsr{r}}}^\star}
\newcommand{\rstartr}{{\tnsr{r}^\star}^\transp}
\newcommand{\xvelgrad}{\tnsr{l}}

\newcommand{\pxvelgradhat}{\hat{\tnsr{l}}^p}
\newcommand{\xdefrate}{\tnsr{d}}
\newcommand{\dxdefrate}{\tnsr{d}^\prime}

\newcommand{\latdefrate}{{\tnsr{d}^{p}}^\prime}
\newcommand{\dlatdefratehat}{\hat{\tnsr{d}^p}^\prime}

\newcommand{\xspin}{\tnsr{w}}
\newcommand{\pxspin}{\tnsr{w}^{p}}
\newcommand{\pxspinhat}{\hat{\tnsr{w}}^{p}}
\newcommand{\spinvec}{\vctr{\omega}}

\newcommand{\lateps}{{\tnsr{e}^e}}
\newcommand{\dlateps}{{\tnsr{e}^e}^\prime}
\newcommand{\latepsdot}{\dot{\tnsr{e}}^e}
\newcommand{\dlatepsdot}{\dot{{\tnsr{e}}^e}^\prime}

\newcommand{\schmid}{{\hat{\vctr{s}}}^{\alpha}\otimes {\hat{\vctr{m}}}^{\alpha}}
\newcommand{\symschmid}{\hat{\tnsr{p}}^\alpha}
\newcommand{\skwschmid}{\hat{\tnsr{q}}^\alpha}
\newcommand{\slppln}{\vctr{m}^\alpha}
\newcommand{\slpdir}{\vctr{s}^\alpha}

\newcommand{\rss}{\tau^\alpha}
\newcommand{\gammadot}{\dot{\gamma}^\alpha}
\newcommand{\gdotz}{\dot{\gamma}_0}
\newcommand{\gdot}{\dot{\gamma}}
\newcommand{\gdots}{\dot{\gamma_s}}
\newcommand{\sumss}{\sum_{\alpha}}
\newcommand{\matlatepsdot}{ \left\{ \dot{ \mathsf{e}^e} \right\} }   
   
\newcommand{\matlateps }{ \left\{\mathsf{e}^e\right\}  }   
\newcommand{\matlatepsold }{ \left\{\mathsf{e}_0^e\right\}  }  
\newcommand{\matdlateps }{ \left\{ {\mathsf{e}^e}^\prime \right\}  }    
\newcommand{\matdlatepsold }{ \left\{ {\mathsf{e}_0^e}^\prime \right\}  }  
\newcommand{\matdefrate }{\Big\{ \mathsf{d} \Big\}   }     
  
\newcommand{\matddefrate }{\Big\{ \mathsf{d}^\prime \Big\}   }
\newcommand{\matlatdefrate }{\Big\{ {\hat{\mathsf{d}}^p} \Big\}   }
\newcommand{\matpxspinhat }{\Big[ {\hat{\mathsf{w}}^p} \Big]   }
  
\newcommand{\matdkirch }{  \left\{ \tau^\prime \right\}  }   
\newcommand{\matkirch }{  \left\{ \tau \right\}  }  
\newcommand{\matdcauchy }{  \left\{ \sigma^\prime \right\}  } 
\newcommand{\matxdelasticity}{\Big[ \,\mathsf{c}^\prime \,\Big]}  
\newcommand{\matxelasticity}{\Big[ \,\mathsf{c} \,\Big]}  
\newcommand{\matxplasticity}{\Big[ \,\mathsf{m} \,\Big]}  
\newcommand{\matxep}{\Big[ \,\mathsf{s} \,\Big]}

\newcommand{\matsymschmid }{ \Big\{\mathsf{p}^\alpha\Big\}  }  
\newcommand{\mathhh}{\Big\{ \mathsf{h} \Big\}   }   
   
\newcommand{\matbodyforce}{\Big\{ \iota \Big\}   } 
 
\newcommand{\mattraction}{\Big\{ t \Big\}   } 

\newcommand{\matvel }{ \Big\{ v \Big\}  } 
\newcommand{\matcapX}{\Big[ \,\mathsf{X} \,\Big]}    
\newcommand{\matcapB}{\Big[ \,\mathsf{B} \,\Big]}
\newcommand{\matcapN}{\Big[ \,\mathsf{N}(\xi, \eta, \zeta) \,\Big]}
\newcommand{\matresidual}{\Big\{ R_u \Big\}}
\newcommand{\matresiduale}{\Big\{ R^{\it ele}_u \Big\}}
\newcommand{\matvelnp}{\Big\{ \mathsf{V} \Big\}}
\newcommand{\delmatvelnp}{\Big\{ \Delta\mathsf{U} \Big\}}
\newcommand{\matstiffd}{\Big[ \,\mathsf{k}^{\it ele}_d \,\Big]} 
\newcommand{\matstiffv}{\Big[ \,\mathsf{k}^{\it ele}_v \,\Big]} 

\newcommand{\Matstiffd}{\Big[ \,\mathsf{K}_d \,\Big]} 
\newcommand{\Matstiffv}{\Big[ \,\mathsf{K}_v \,\Big]} 
\newcommand{\matdelta}{\Big\{ \mathsf{\delta} \Big\}}

\newcommand{\surfnorm}{{\vctr{\nu}(\vctr{x})}}

\newcommand{\dee}{{\mathrm{d}}}

\begin{document}
%
% Title page.
%
\title{ \fepx\, -- Finite Element Polycrystals\\
Theory,  Finite Element Formulation,  Numerical Implementation\\
 and Illustrative Examples\\
}
\author{Paul R. Dawson and Donald E. Boyce\\
Deformation Process Laboratory\footnote{http://anisotropy.mae.cornell.edu/dplab/}\\
Sibley School of Mechanical and Aerospace Engineering\\
 Cornell University
 }
\date{\today}

\maketitle

\begin{figure}[h!]
\centering
{
\includegraphics*[width=10cm]{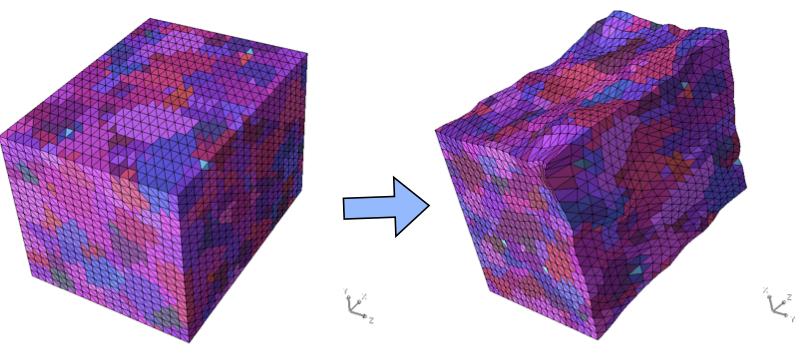}
}\\
Virtual Polycrystal Deformed by \fepx
\end{figure}

%
% Abstract
%
\begin{abstract}
\fepx\, is a modeling framework  for computing the elastoplastic deformations of polycrystalline solids.  
Using the framework, one can simulate the mechanical behavior of aggregates of crystals, referred to as virtual polycrystals, over large strain deformation paths.  This article presents the theory, the finite element formulation, and important features of the numerical implementation that collectively define the modeling framework.    The article also provides several examples of  simulating the elastoplastic behavior of polycrystalline solids to illustrate possible applications of the framework.  There is an associated finite element code, also referred to as \fepx, that is based on the framework presented here and was used to perform the simulations presented in the examples.  The article serves as a citable reference for the modeling framework for users of that code.  Specific information about the formats of the input and output data, the code architecture, and the code archive are contained in other documents.  

\end{abstract}
\pagebreak[4]

\tableofcontents\pagebreak[4]
\listoffigures\pagebreak[4]
\section{Introduction}
\label{chapter:purpose}
\subsection{Purpose }
\label{sec:purpose}
The purpose of this article is to lay out a complete system of equations for modeling the anisotropic, elasto-viscoplastic response of polycrystalline solids comprised of aggregates of grains and to summarize a finite element formulation that may be  employed to compute the motion and stress in polycrystals governed by the system of equations under imposed loadings.  The governing equations together with associated solution methodologies define a modeling framework, referred to as \fepx, that is focused at a physical length scale of an ensemble of grains.   There is an associated finite element code, also named \fepx, that follows the framework.   A major motivation for archiving this article is to provide a thorough and accessible reference  that  researchers who utilize the code can readily cite.  However, the article 
stands independently in providing a complete summary of a crystal-scale model for the elasto-viscoplastic response of polycrystalline aggregates and a finite element formulation that enables solving the model equations over motions that entail large deformations.

The content provided here regarding the governing equations and finite element framework draws primarily from  the following published articles: 
\cite{daw_mar_98p,mar_daw_98a,mar_daw_98b,bar_daw_mil_99}.
The present article is not intended to serve as a primer for 
computational crystal plasticity, so  background knowledge of solid mechanics, including crystal plasticity,
 and nonlinear finite element methods  is assumed.   Rather, it strives to encapsulate the full set of equations, assumptions, and solution approximations necessary to document simulation results with sufficient detail to
 facilitate those results being reproduced by others.

\subsection{Scope}
\label{sec:scope}  
The scope of this article is limited to the theory and methods that define the \fepx\, framework, plus a general overview of
the data flow within the framework and the interfaces with tools to instantiate virtual polycrystals and to
visualize simulation results.   
Consequently, there are sections of the article devoted
to these topics, as listed in the Table of Contents.  Also provided are representative examples to illustrate application
of the derivative finite element code to modeling of single and dual phase metallic alloys.
No detailed information is included on the specific formatting used for problem definition, code execution, or exported simulation results.
That information is contained in separate documentation associated with the code itself.      

\subsection{Complementary modeling tools}
\label{sec:complementarysoftwar}
The role of \fepx\, in the modeling of polycrystals is to solve the boundary value problem associated with the elastoplastic 
response of a polycrystalline solid arising from applied mechanical loading.   Separate tools are needed to instantiate a virtual polycrystal and 
to discretize it with finite elements.   \fepx\, accepts the finite element mesh generated by custom \matlab\, scripts (available in the \odfpf\, package) and by the \neper\, program~\cite{que_daw_bar_11}.   Separate packages for visualizations also are needed.  Export scripts are available for writing files that can be  imported by \paraview\cite{paraview}, and \visit\cite{HPV:VisIt}.

\pagebreak[4]
\section{Capabilities of \fepx}
\label{chapter:capabilities}
The \fepx\, framework is a combination of a theoretical construct developed by a world-wide community of researchers and a numerical solution methodology (finite element formulation) developed by the DPLab members over the past than two decades. 
The governing equations and solution methodology become tightly intwined through the choice of interpolation functions for the motion and the weighted residuals for equilibrium.   Thus, we refer to the combination as a framework and do not make a strong distinction between the framework and the derivative code, calling both \fepx.   

\subsection{What \fepx\, can do.}
\label{sec:candu}
\fepx\, is useful for simulating the mechanical behavior of polycrystalline solids at the level of aggregates of grains.  
The aggregates may be comprised of grains of a single phase or of multiple phases.
Grains are discretized with finite elements so any sub-volume of an element is a sub-volume of an individual crystal.
The local behaviors associated with the material with any element correspondingly are those of a crystal.  
In particular, the behaviors include:
\begin{itemize}
\item{nonlinear kinematics capable of handling motions with both large strains and large rotations;}
\item{anisotropic elasticity based on cubic or hexagonal crystal symmetry;   }
\item{anisotropic plasticity based on rate-dependent slip on a restricted number of systems for cubic or hexagonal symmetry; } 
\item{evolution of state variables for crystal lattice orientation and slip system strengths;} 
\end{itemize}
To accommodate these behaviors the finite element formulation has incorporated a number a numerical features, such as:
\begin{itemize}
\item{higher-order, isoparametric elements with quadrature for integrating over the volume;} 
\item{implicit update of the stress in integrations over time;}
\item{monotonic and cyclic loading under quasi-static conditions;}
\end{itemize}

Depending on the goals of the simulation,  aggregates might number in grains from only a few to tens of thousands (or more).   
The grains can be discretized at a level appropriate for the intent of a simulation.  
The number of grains together with the level of discretization within grains set the computational burden for a simulation.  
To accommodate combinations with high burden, the \fepx\, code employs scalable parallel methods and executes on clusters.
\fepx\, has been developed to use meshes constructed by instantiation tools for virtual polycrystals and to output data in 
an archivable format for subsequent use with visualization tools or other interpretation tools.

With these capabilities \fepx\, is well-suited, for example,  to model the mechanical behavior of polycrystals that exhibit inhomogeneous deformations within and among the crystals,  to investigate the heterogeneity of stress within a polycrystal,  or to examine the roles of neighbors on the behaviors of individual grains.  When teamed with appropriate instantiation methods, \fepx\, can be used effectively to model yielding and flow of alloys with complicated phase/grain topologies and morphologies.  
 
\subsection{What \fepx\, cannot do.}
\label{sec:nocandu}
The \fepx\, framework does not encompass many aspects of the behaviors observed in real materials.   Some of its principal limitations include:
\begin{itemize}
\item{plastic flow occurs by slip  -- no other mechanisms, such as twinning and creep, are modeled; }
\item{deformations are ductile  -- no fracture models are included;}
\item{loading is quasi-static -- no inertial effects are modeled; }
\item{loading is mechanical (isothermal) --  coupling with heat transfer (or other physical processes) is not considered;}
\item{boundary conditions are simple  --  neither friction models nor changing contact conditions are included.}
\end{itemize}
With these limitations,  \fepx\, is not well-suited for modeling applications with complex loading conditions, such as many metal forming  and joining processes, or for modeling applications involving fragmentation failure of a dynamically loaded body.

\subsection{What \fepx\, has been used to study.}
\label{sec:our_studies}
A variety of interesting problems arise at physical length scales in which a sample volume encompasses an aggregate of grains.  We typically think of an aggregate containing on the order of $10^3-10^6$ grains, but the simulation framework embodied in \fepx\, is appropriate for single-grain or multi-grain samples ($1-10^2$ grains), as well.   Some of the applications of \fepx\, published in the open literature are listed below.   Users of \fepx\, are encouraged to examine articles in areas of interest to obtain information beyond the scope of this article resulting from the  collective experiences of others in applying \fepx. 
\begin{itemize}
\item{Grain interactions with attention focused on bulk texture evolution.
Articles published in this area 
are: \cite{bea_daw_mat_koc_95,mar_daw_jen_95,sar_daw_96a,sar_daw_96b,mik_daw_98,leb_daw_ker_wen_03,mir_daw_lef_07}.}
\item{Deformation heterogeneity within the grains comprising an aggregate with focus on intra-grain misorientation distributions. Articles published in this area 
are: \cite{mik_daw_99,bar_daw_01b,bar_daw_02,que_daw_dri_12}.}
\item{Inter- and intra-grain stress/elastic strain distributions, especially including comparisons to neutron and x-ray diffraction experiments.
Articles published in this area 
are: \cite{daw_boy_mac_rog_00,daw_boy_mac_rog_01,daw_boy_rog_05b,mil_par_daw_han_08,rit_daw_mar_10,mar_daw_gar_12}.}
\item{The elasto-plastic transition occurring during loading of polycrystalline solids, with focus on the redirection of stress at the grain level. 
Articles published in this area are: \cite{bar_daw_mil_99,han_daw_05b,won_daw_10,sch_won_daw_mil_13}}.
\item{Cyclic loading with interest in evolution of stress and its implications for fatigue failure.
Articles published in this area are: \cite{tur_log_daw_mil_03,won_daw_11}.}
\item{Evolution of dislocation density and associated peak broadening.
Articles published in this area are: \cite{daw_boy_rog_05,won_par_mil_daw_13}}.
\item{Virtual polycrystal instantiation issues, including sensitivity of the stress and deformation to discretization.  Articles published in this area are: \cite{log_tur_mil_rog_daw_04,rit_daw_09, que_daw_bar_11}.}
\end{itemize}

Much of the original work  in utilizing polycrystal plasticity constitutive models within finite element simulations was focused on bulk texture evolution in macroscopic scale deformation
processes, such as metal forming operations (rolling, extrusion, and sheet forming) and geological processes (particularly mantle convection).  In such cases, the relative sizes of finite elements and grains  were reversed in comparison to those of the intended applications of the \fepx\, framework described in this article.  Consequently, the mechanical properties within an element were derived from an average over an ensemble of crystals once an averaging hypothesis ({\it e.g.} isostress or isostrain) was imposed.  
Examples of this type of application are: \cite{mat_daw_89,mat_daw_koc_90,bea_mat_daw_joh_93,bea_daw_mat_koc_kor_94,kum_daw_95,daw_wen_00,daw_mac_wu_03,daw_boy_rog_05b}.    While relevant from a historical perspective in the development of \fepx, the  
\fepx\, framework described here does not include evaluating properties within an element on the basis an average over a population of grains.  Rather, properties within an element are those of a single orientation, consistent with an element spanning only a part of any given grain.

\pagebreak[4]
\section{Nomenclature}
\label{chapter:nomenclature}

\subsection{Variables used in theoretical description} 
\label{sec:nomen-theory}

\begin{tabular}[b]{lll}

\nomenclature{$\beta$}{determinant of the elastic stretch, $\latstretch$ }\\

\nomenclature{$\gammadot$}{shearing rate of the $\alpha$ slip system }\\

\nomenclature{$\vctr{\nu}$}{surface normal vector}\\

\nomenclature{$\vctr{\iota}$}{body force vector}\\

\nomenclature{$\pi$}{mean stress (${\rm tr}(\cauchy)/3$) }\\

\nomenclature{$\phi$}{rotation angle associated with $\rodvec$}\\

\nomenclature{$\cauchy$}{Cauchy stress }\\

\nomenclature{$\kirch$}{Kirchhoff stress }\\

\nomenclature{$\rss$}{resolved shear stress on the $\alpha$ slip system }\\

\nomenclature{$\spinvec$}{spin vector associated with $\rodvel$ }\\

\nomenclature{$\vctr{\chi}$}{mapping function of motion}\\

\nomenclature{$\cal{B}$}{domain of the polycrystal}\\

\nomenclature{${\partial {\cal B}}$}{surface of the polycrystal}\\

\nomenclature{$\boldsymbol{\cal{C}}$}{elasticity (stiffness) tensor}\\

\nomenclature{$\tnsr{d}$}{deformation rate (symmetric part of $\tnsr{l}$)}\\

\nomenclature{$\dlatdefratehat$}{plastic deformation rate (symmetric part of $\hat{\tnsr{l}}^p$)}\\

\nomenclature{$\lateps$}{elastic strain}\\

\nomenclature{$\tnsr{f}$}{deformation gradient}\\

\nomenclature{$\tnsr{f}^e$}{elastic part of the deformation gradient}\\

\nomenclature{$\tnsr{f}^\star$}{rotational part of the deformation gradient associated with the lattice rotation}\\

\nomenclature{$\tnsr{f}^p$}{plastic part of the deformation gradient}\\

\nomenclature{$g^\alpha$}{ strength of  the $\alpha$ slip system}\\

\nomenclature{$\tnsr{l}$}{velocity gradient}\\

\nomenclature{$\hat{\tnsr{l}}^p$}{plastic velocity gradient}\\

\nomenclature{$\slppln$}{normal to the slip plane for the $\alpha$ slip system}\\

\nomenclature{$\boldsymbol{\cal{M}}$}{plasticity (stiffness) tensor}\\

\nomenclature{$\vctr{n}$}{axis vector associated with $\rodvec$}\\

\nomenclature{$\tnsr{p}^\alpha$}{symmetric part of the Schmid tensor for the $\alpha$ slip system }\\

\nomenclature{$\tnsr{q}^\alpha$}{skew part of the Schmid tensor  for the $\alpha$ slip system}\\

\nomenclature{$\quaternion$}{quaternion representation of lattice orientation $(q_0, \vec{q})$}\\

\nomenclature{$\rodvec$}{Rodrigues vector for the orientation of the crystallographic lattice}\\

\nomenclature{$\rstar$}{rotational part of the deformation gradient associated with the lattice rotation ($=\tnsr{f}^\star $)}\\

\nomenclature{$\tnsr{R}$}{rotation operator corresponding to $\rodvec$}\\

\nomenclature{$\slpdir$}{slip direction for the $\alpha$ slip system}\\

\nomenclature{$\vctr{t}$}{traction vector}\\

\nomenclature{$\bar{\vctr{t}}$}{imposed traction vector on the surface}\\

\nomenclature{$\Delta t$}{time step}\\

\nomenclature{$\vctr{v}$}{velocity vector of a point in the current configuration}\\

\nomenclature{$\bar{\vctr{v}}$}{ imposed velocity vector on the surface}\\

\nomenclature{$\latstretch$}{elastic stretch}\\

\nomenclature{$\tnsr{w}$}{spin (skew part of $\tnsr{l}$)}\\

\nomenclature{$\pxspinhat$}{plastic spin (skew part of $\hat{\tnsr{l}}^p$)}\\

\nomenclature{$\vctr{x}$}{position vector of a point in the current configuration}\\

\nomenclature{$\vctr{X}$}{position vector of a point in the reference configuration}
\end{tabular}
\bigskip

\subsection{Parameters appearing in the constitutive models} 
\label{sec:nomen-parameters}

\begin{tabular}[b]{lll}

\nomenclature{$\gdotz$}{fixed-state strain rate scaling coefficient }\\

\nomenclature{$\gdots$}{saturation strength strain rate scaling coefficient}\\

\nomenclature{$\kappa$}{elastic bulk modulus}\\

\nomenclature{$c_{ij}$}{components of the elastic stiffness}\\

\nomenclature{$g_{0}$}{initial slip system strength}\\

\nomenclature{$g_{1}$}{reference value of saturation strength}\\

\nomenclature{$h_{0}$}{strength hardening rate coefficient }\\

\nomenclature{$m$}{fixed-state strain rate sensitivity}\\

\nomenclature{$m^\prime$}{saturation strength rate scaling exponent}\\

\nomenclature{$n^\prime$}{power on modified Voce hardening term }

\end{tabular}\bigskip

\subsection{Variables used in implementation description} 
\label{sec:nomen-implementation}

\begin{tabular}[b]{lll}

\nomenclature{$ \left\{ \delta \right\}$}{matrix trace operator} \\

\nomenclature{$\Big\{ \psi \Big\}$}{matrix form of the weights}\\

\nomenclature{$ \Big\{ {\it \Psi } \Big\}$}{nodal point weights vector  }\\

\nomenclature{$\left\{ \sigma \right\}$}{vector matrix form of Cauchy stress}\\

\nomenclature{$\left\{ \tau \right\}$}{vector matrix form of Kirchhoff stress}\\

\nomenclature{$ \matcapB $}{spatial derivatives of the interpolation functions, $[ N ]$  }\\

\nomenclature{$\matxelasticity$}{matrix form of the elastic stiffness}\\

\nomenclature{$\left\{ \sf{d} \right\}$}{vector matrix form of deformation rate}\\

\nomenclature{$\left\{ \sf{e}^e \right\}$}{vector matrix form of elastic strain}\\

\nomenclature{$\Big\{ \mathsf{f}^e_a \Big\} , \Big\{ \mathsf{F}_a \Big\}$}{elemental and global surface traction and body force matrices}\\

\nomenclature{$\Big\{ \mathsf{f}^e_v \Big\} , \Big\{ \mathsf{F}_d \Big\}$}{elemental and global initial elastic strain matrices}\\

\nomenclature{$\Big\{ \mathsf{f}^e_d \Big\} , \Big\{ \mathsf{F}_v \Big\}$}{elemental and global spin correction stiffness matrices}\\

\nomenclature{$\mathhh$}{vector matrix form of the spin correction and initial elastic strain terms  }\\

\nomenclature{$\matstiffd, \Matstiffd$}{elemental and global deviatoric stiffness matrices}\\

\nomenclature{$\matstiffv, \Matstiffv$}{elemental and global volumetric stiffness matrices}\\

\nomenclature{$\matxplasticity$}{matrix form of the plastic stiffness}\\

\nomenclature{$[ N ]$}{interpolation functions for interpolation of the velocity distribution}\\

\nomenclature{$\matsymschmid$}{vector matrix form of the symmetric part of the Schmid tensor}\\

\nomenclature{$R_u$}{equilibrium weighted residual  }\\

\nomenclature{$\matresidual$}{elemental residual vector}\\

\nomenclature{$\matpxspinhat $}{matrix form of the plastic spin }\\

\nomenclature{$\matvel $}{matrix form of the velocity}\\

\nomenclature{$ \matvelnp $}{nodal point velocity vector  }\\

\nomenclature{$\left\{ x \right\}$}{matrix form of the position}\\

\nomenclature{$\matcapX$}{coefficient matrix of factors necessary to deliver correct inner product from matrix multiplications.}

\end{tabular}

%\nomenclature{$-$}{-}%
%\nomenclature{$-$}{-}%
%\nomenclature{$-$}{-}%
%\nomenclature{$-$}{-}%
%\nomenclature{$-$}{-}%

\pagebreak[4]
\section{Governing Equations}
\label{chapter:governing_equations}
\begin{figure}[h]
\begin{center}
\includegraphics*[width=10cm]{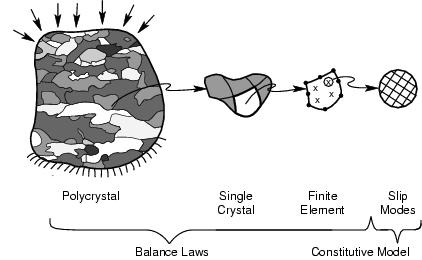}
\caption{Schematic diagram of a polycrystal subjected to mechanical loading and the discretization of 
individual crystals with finite elements.}
\label{fig:small_scale}
\end{center}
\end{figure}

The schematic diagram shown in Figure~\ref{fig:small_scale} is a useful depiction of the intended application of the \fepx.  
The body is a polycrystal in which the full volume is subdivided into the individual crystals such that the 
entire volume is completely filled.  The inter-crystal boundaries are cohesive so that there is neither sliding 
at the crystal boundaries nor separation between crystals.  Every crystal is discretized with one or more
finite elements.   The material properties are evaluated at quadrature points within the elements 
on the basis of single crystal behavior, which is intimately tied to the crystal structure and its orientation
with respect to the body at large.  External load are applied to the polycrystal along with the
appropriate kinematic constraints.

\subsection{Kinematics and balance laws}
\label{sec:strongforms}
The motion of the virtual polycrystal is assumed to be a smooth mapping of all points within
the domain, $\cal{B}$.  The domain in this context refers to the union of all crystal volumes that collectively 
define the virtual polycrystal.  The internal crystal interfaces are smooth surfaces that remain
coherent throughout the motion.   Under the motion, the current coordinates may be written as
a function of  reference coordinates for every point in the domain:
\begin{equation}
\vctr{x} = \vctr{\chi}(\vctr{X})
\label{eq:mapping}
\end{equation}
The deformation gradient is defined locally as:
\begin{equation}
\tnsr{f} = \frac{\partial\vctr{x}}{\partial\vctr{X}}
\label{eq:defgrad}
\end{equation}
While the mapping, $\chi$, is smooth, the deformation gradient will be so only within the interior of elements.  
Discontinuities can arise across element boundaries, whether the boundaries lie within crystals or on the interface between crystals.
The time-rate-change of the deformation gradient is given by:
\begin{equation}
\dot{\tnsr{f}} = \tnsr{l} \cdot \tnsr{f} 
\end{equation}
where the velocity gradient, $\tnsr{l}$, is computed from the velocity field, $\vctr{v}(\vctr{x})$, as: 
\begin{equation}
\tnsr{l} = \frac{\partial\vctr{v}}{\partial\vctr{x}}
\label{eq:velocitygradient}
\end{equation}
Again, while the velocity is smooth everywhere, its gradient may have discontinuities across element boundaries.  
The velocity gradient is decomposed into its symmetric and skew parts:
\begin{equation}
\tnsr{l} = \tnsr{d} +\tnsr{w}
\label{eq:velgrad_decomp}
\end{equation}
where $\tnsr{d}$ is the deformation rate (symmetric part) and $\tnsr{w}$ is the spin (skew part).  

The motion of the polycrystal is driven by the stress.   
The Cauchy stress, $\cauchy(\vctr{x})$, is a field variable defined over the polycrystal domain.
Under the loading assumptions, inertia in the balance of linear momentum is neglected,
giving static equilibrium in the local form as:
\begin{equation}
\divop {\cauchy}^T + \vctr{\iota} \,=\, \boldsymbol{0}
\label{eq:equilibrium}
\end{equation}
where $\vctr{\iota}$ is the body force vector.  Body forces are neglected in the current implementation of \fepx.
This equation applies to the interior of the crystals.   Across crystal interfaces, continuity of the
traction is needed.  Applying the Cauchy formula, this condition may be written
for two contacting crystals, $i$ and $j$, as:
\begin{equation}
\vctr{t}(\vctr{x})^i  \,=\,    \ \vctr{t}(\vctr{x})^j
\label{eq:crystal_tract_vec}
\end{equation}
where the tractions are related to the stress by means of the Cauchy formula:
\begin{equation}
\vctr{t} \,=\,    \surfnorm \cdot  \cauchy 
\label{eq:cauchyformula}
\end{equation}
The Cauchy stress may be split into deviatoric and spherical parts:
\begin{equation}
\cauchy = \dcauchy - \pi \tnsr{ I}
\label{eq:deviator}
\end{equation}
which is central to the material response as only the deviatoric part drives plastic flow.

\subsection{Constitutive equations}
\label{sec:constitutive}
The material behavior is quantified with a set of constitutive equations, here written at the level of the single crystal.
The behavior includes both elastic (recoverable strains upon removal of the stress) and plastic (non-recoverable strains upon removal of the stress) responses.   
These can occur concurrently which requires coupling of the motions for the two via a kinematic decomposition.   The decomposition is not strictly derivable from the mapping, but rather involve assumptions regarding the behavior.  Consequently, it is part of constitutive model.    The elastic response is limited to linear behavior following Hooke's law for anisotropic behavior.  The plastic response is nonlinear and rate-dependent (viscoplastic).   It is assumed to be isochoric and independent of the mean stress.   The set of equations are summarized in the following subsections, starting with the kinematic decomposition.  The equations for the elastic and plastic responses follow discussion of the kinematic decomposition and are broken into two parts:  fixed state relations and evolution relations.  
\subsubsection{Elastoplastic kinematic decomposition}
\begin{figure}[h]
\begin{center}
\includegraphics*[width=8cm]{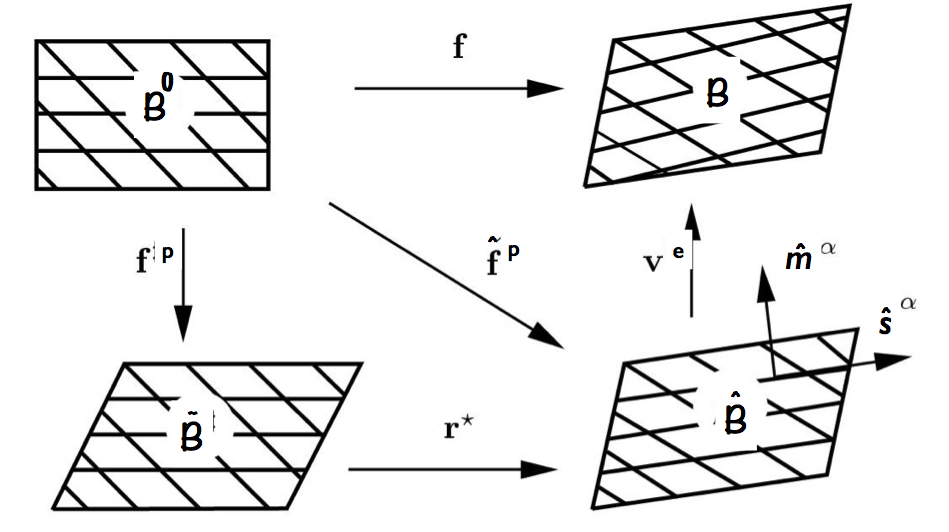}
\caption{Kinematic decomposition for motion by a combination of plastic slip, rotation and elastic straining.   }
\label{fig:kinematic_decomposition}
\end{center}
\end{figure}

The  deformation at a material point\footnote{In the context of \fepx, a material point is a volume of material that is small in comparison to the finite element in which it resides (and thus, small in comparison to an individual crystal), yet large enough to fully reflect the crystal structure and the deformation processes.  For slip, this means that the dimensions of the volume are much larger than a Burger's vector.}
is  a combination of the elastic and plastic parts.   In addition, a rotation occurs as part of the complete motion.  
 These are shown schematically in Figure~\ref{fig:kinematic_decomposition}. 
The decomposition that describes this motion consists of breaking the deformation gradient into a sequence of three parts: a plastic part, a rotation and an elastic part, given as:
\begin{equation}
 \xdefgrad = \tnsr{f}^e \tnsr{f}^\star \tnsr{f}^p = \latstretch \rstar\pxdefgrad
 \label{eq:kinematic_decomp}
\end{equation}
Each part of the decomposition brings the material point to a new configuration, starting with reference coordinates, $\vctr{X}$, and finishing at the current coordinates, $\vctr{x}$.
The elastic part is a pure stretch which,  by assuming small elastic strains,  can be approximated with:
\begin{equation}
\latstretch = \tnsr{I} + \lateps
\label{eq:small_elastic_stretch}
\end{equation}
where
\begin{equation}
|| \lateps || << 1 
\end{equation}
The plastic part involves both stretch and rotation as a consequence of being a linear combination
of slip modes, each of which is simple shear.     The distinct rotation in $\tnsr{f}^\star$ (or equivalently, $\rstar$) 
is the rotation beyond that included in $\pxdefgrad$ that is needed for consistency with the overall mapping 
given by Equation~\ref{eq:mapping}.

The primitive solution field variable of \fepx\, is the velocity, owing principally to the code's legacy of being a tool for  modeling
viscoplastic flow.  To cast the kinematic decomposition in rate form, the velocity gradient first is written in terms of the deformation gradient and its time-rate-of-change:
\begin{equation}
\xvelgrad = \xdefgraddot \xdefgradi 
\label{eq:kinematic_decomp_rate}
\end{equation}
where the velocity gradient is subsequently decomposed into the deformation rate and spin, as per Equation~\ref{eq:velgrad_decomp}.  The deviatoric deformation rate is obtained by subtracting the volumetric part from the total:  
\begin{equation}
\dxdefrate = \xdefrate - \frac{1}{3}\trace{\xdefrate} 
\label{eq:deviatoric_defrate}
\end{equation}
Substituting Equation~\ref{eq:kinematic_decomp} into Equation~\ref{eq:kinematic_decomp_rate} and separating  the deformation rate into its volumetric and deviatioric parts with Equation~\ref{eq:deviatoric_defrate} gives:
\begin{equation}
\trace ( \xdefrate) = \trace( \latepsdot )
\label{eq:kinematic_decomp_meandefrate}
\end{equation}
and
\begin{equation}
\dxdefrate = \dlatepsdot +  \dlatdefratehat +
\dlateps \pxspinhat - \pxspinhat \dlateps
\label{eq:kinematic_decomp_devdefrate}
\end{equation}
in which the small elastic strain approximation from Equation~\ref{eq:small_elastic_stretch} has been invoked.  
The spin becomes:
\begin{equation}
\xspin = \pxspinhat + \dlateps \dlatdefratehat - \dlatdefratehat \dlateps 
\label{eq:kinematic_decomp_spin}
\end{equation}
where, again, small elastic strains are assumed.  
In Equations~\ref{eq:kinematic_decomp_spin} and \ref{eq:kinematic_decomp_devdefrate}, 
the hat over $\pxspin$ and $ \latdefrate$ indicates mapping to configuration $\cal{\hat B}$ using $\rstar$, as indicated in Figure~\ref{fig:kinematic_decomposition}:
\begin{equation}
 \pxspinhat = \rstar \pxspin {\rstar}^\transp
\end{equation}
\begin{equation}
 \dlatdefratehat = \rstar \latdefrate {\rstar}^\transp
\end{equation}
The deformations associated with elastic and plastic parts of the kinematic decomposition are intimately connected to the 
crystallographic lattice.  The orientation of the lattice relative to a set of global base vectors  at a designated material point is given by  
the associated quaternion, $\quaternion$, which is parameterized by the components:  $(q_0, \vec{q})$.
It is convenient to use other representations for orientation as well, depending on the task at hand.  
Two frequently used representations are the Rodrigues vector:
\begin{equation}
        \vctr{r} = \vctr{n} \tan \frac{\phi}{2} = \vec{q}/q_0
        \label{eq:rod_vect}
\end{equation}
and the rotation tensor:
\begin{equation}
        \tnsr{R} = {\frac{1}{1+\vctr{r} \cdot \vctr{r}}}\big({{\tnsr{I}(1-\vctr{r} \cdot \vctr{r})}+{2(\vctr{r} \otimes \vctr{r} + \tnsr{I}\times \vctr{r})}}\big)
        \label{eq:rotation_tensor}
\end{equation}
Properties that are dependent on lattice orientation are indicated as a function of $\quaternion$.  Note also that
changes in the lattice orientation that accompany the motion are embedded in $\rstar$, so the evolution rate of 
$\quaternion$ is defined in terms of the rate of change of $\rstar$.  
With the kinematic decomposition specified, the Kirchhoff stress is written based on
the material point volume in the $\hat{\cal B}$ configuration as:
\begin{equation}
\kirch = \beta \cauchy  
\hspace{0.5 cm} {\rm where} \hspace{0.5cm} 
\beta = \detop (\latstretch) 
\label{eq:kirchhoff_defn}
\end{equation}

\subsubsection{Fixed state constitutive relations }

The kinematic decomposition must be accompanied by equations relating the elastic and plastic deformations to the stress.
For the elastic deformations,  this relation is simply Hooke's law, written using the $\hat{\cal B}$ configuration as a reference volume:
\begin{equation}
\kirch = \cofq \lateps 
\label{eq:hookes_law}
\end{equation}
Here, the anisotropic behavior stemming from the crystal symmetry is indicated by the 
orientation dependence of the elastic stiffness.
The structure of the elastic stiffness (occurrence of zero or repeated components in $\cofq$) reflects the application of symmetry conditions to the fully anisotropic version of Hooke's law.  These are more easily presented using a vector representation of
the stress and strain tensors, as is done in Section~\ref{chapter:fem_implementation}.  

\begin{figure}[h]
\centering
\subfigure[FCC slip systems ]
{
 \includegraphics*[width=5cm]{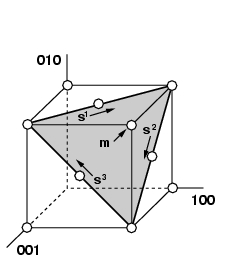}
}
\vspace{1cm}
\subfigure[BCC slip systems]
{
   \includegraphics*[width=6.6cm]{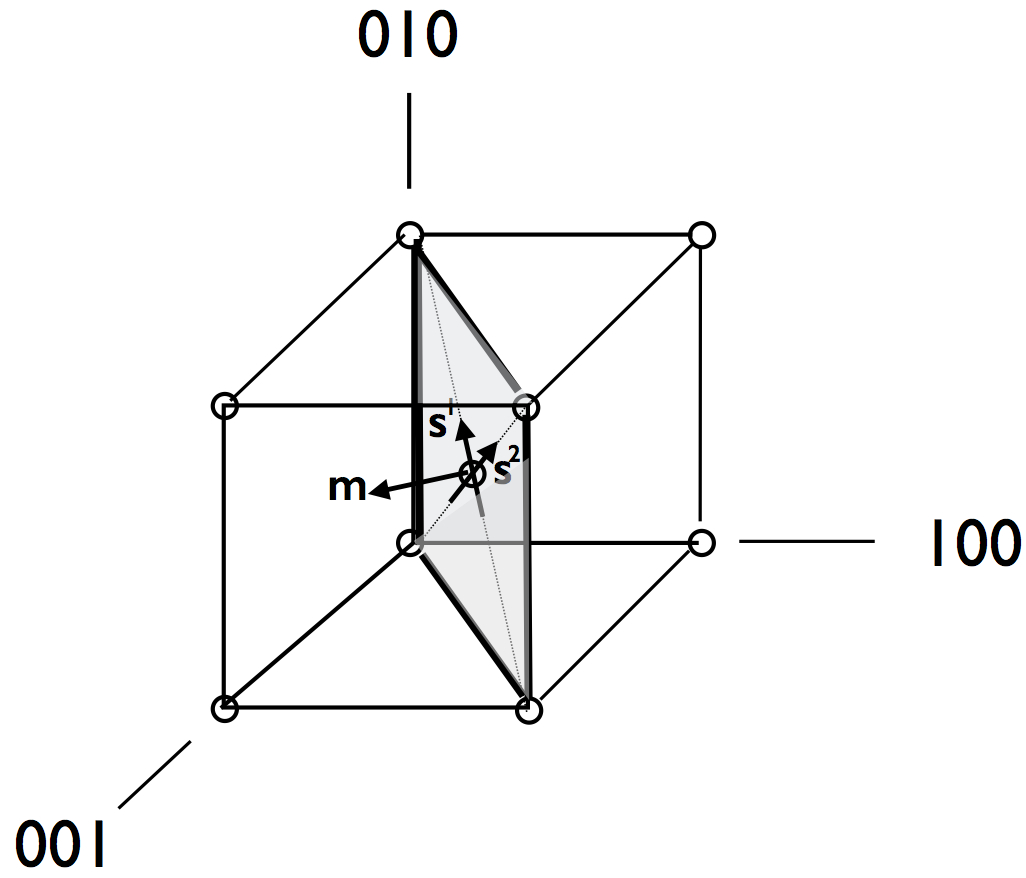}
}
\caption{Primary slip systems for face-centered cubic (FCC) and body-centered cubic (BCC) crystal types.}
\label{fig:fcc_ss}
\end{figure}

\begin{figure}[h]
\begin{center}
\includegraphics*[width=12cm]{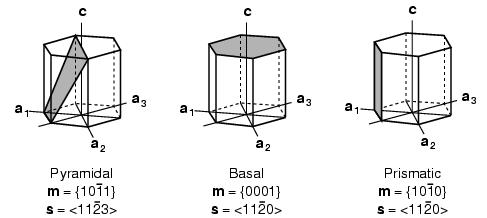}
\caption{Primary slip systems for hexagonal close-packed (HCP) crystal types.}
\label{fig:hcp_ss}
\end{center}
\end{figure}

For the plastic flow, the relation is a combination of several equations that describe crystallographic slip on a limited number of slip systems (commonly called restricted slip).  First are equations for the kinematic decomposition written in terms of slip using
the Schmid tensor's symmetric and skew parts:  
\begin{equation}
\pxvelgradhat 
=\dlatdefratehat + \pxspinhat
\label{eq:plastic_velgrad}
\end{equation}
where
\begin{equation}
\dlatdefratehat =  \sumss \gammadot \symschmid
\hspace{0.5 cm} {\rm and} \hspace{0.5cm}
\pxspinhat =  \rstardot \rstartr +  \sumss \gammadot \skwschmid
\label{eq:ss_superposition}
\end{equation}
and 
\begin{equation}
\symschmid  =  \symschmid(\quaternion)  = \mbox{\rm sym} \, (\schmid)
\end{equation}
\begin{equation}
\skwschmid  =   \skwschmid(\quaternion)  = \mbox{\rm skw} \, (\schmid)
\end{equation}
The slip systems commonly assumed for face-centered cubic and hexagonal close-packed crystal types
are shown in Figures~\ref{fig:fcc_ss} and \ref{fig:hcp_ss}, respectively. 
\begin{table}[h]
\centering
\begin{tabular}{||c|c|c|c|c||}
\hline
\hline
Crystal Type & Name & Number & $\vctr{m}$ & $\vctr{s}$ \\
\hline
FCC  & octahedral                & 12 & $\{111\} $& $<110> $\\
BCC &  -               & 12& $\{110\} $& $<111> $\\
HCP & basal        & 6  &  $\{0001\} $& $<11\bar{2}0> $\\
- & prismatic  & 6  &  $\{10\bar{1}0\} $& $<1 1 \bar{2}0> $\\
- & pyramidal & 6  &  $\{10\bar{1}1\} $& $<1 1 \bar{2}3> $\\
\hline
\hline
\end{tabular}
\caption{Slip system vectors for (unstressed) FCC, BCC and HCP crystal types given a coordinate system attached to the lattice orientation.
}
\label{tab:slip_systems}
\end{table}

Next is an equation that defines the kinetics of slip, which introduces the rate dependence of plastic flow
using a power law expression between the resolved shear stress and the slip system shearing rate:
\begin{equation}
\gammadot = f(\rss , g^\alpha) = \gdotz \left( {\frac{ \arrowvert \rss \arrowvert }{g^\alpha}}\right)^{\frac{1}{m} }{\rm sgn}(\rss) 
\label{eq:ss_kinetics}
\end{equation}
The resolved shear stress is scaled by the slip system strength, $g^\alpha$, which in general may be different
for each slip system.   However, in \fepx\, the slip system strengths are the same for each family of slip systems
within a grain.   Thus, the slip system strengths are all the same within each of the finite elements that
discretize a grain for either  FCC and BCC crystal types.
For HCP crystals, the basal, prismatic and pyramidal strengths within each of the finite elements discretizing a grain can have different values, but all the systems
of a given type have the same value.     The values of the strength evolve with deformation according to 
the evolution equations, as discussed in Section~\ref{sec:state_evolution}.   The `sgn' term forces the shearing 
to be in the same direction as the shear stress.  Finally, the resolved shear stress is the projection of  the 
crystal stress tensor onto the slip plane and into the slip direction, which is readily computed with the 
symmetric part of the Schmid tensor:
\begin{equation}
\rss=\trace( \symschmid  \dkirch)
\label{eq:rss_projection}
\end{equation}
The equations for slip are combined in a single, nonlinear relation as:
\begin{equation}
\dlatdefratehat  = \mofq \dkirch
\label{eq:ss_const_law}
\end{equation}
 Combining Equations~\ref{eq:kinematic_decomp_meandefrate}, \ref{eq:kinematic_decomp_devdefrate},  \ref{eq:hookes_law} and \ref{eq:ss_const_law} into a single equation that relates the Cauchy stress to the
 total deformation rate requires an additional step to discretize the elastic strain rate.   This step is
 introduced in Section~\ref{chapter:fem_implementation}.

%\parbox[t]{10cm}{
%\begin{center}
% {\bf  ADD:  Elaborate on how we handle the different slip system strengths in the hexagonal case.}
%\end{center}
%}

\subsubsection{State evolution equations}
\label{sec:state_evolution}
There are two state variables at every material point that are updated as a deformation progresses,
the lattice orientation and the slip system strength (also called hardnesses)\footnote{One could argue that the elastic strain also is a state variable, as it quantifies the shape of the unit cell.  However, it is updated  as an integral part of solving for the velocity field, rather than separate from solving for the velocity field as are summarized in this section for the lattice orientations and slip system hardnesses.  See Section~\ref{chapter:fem_implementation}.}.  The rate of lattice re-orientation follows directly from
the Equations~\ref{eq:plastic_velgrad} and \ref{eq:ss_superposition}, assuming that the slip system shearing rates are known.  
Written in terms of the Rodrigues vector:
\begin{equation}
\dot\rodvec = \frac{1}{2} \spinvec + (\spinvec \cdot \rodvec) \rodvec + \spinvec \times \rodvec
\end{equation}
where
\begin{equation}
\spinvec = \vectop\left( \pxspinhat - \sumss \gammadot \skwschmid \right)
\end{equation}

Evolution of the slip system strengths is governed by an additional, empirical relationship which follows
as modified Voce form:
\begin{equation}
\dot{g^\alpha} = h_{0} {\left(\frac{g_{s}(\gdot) - g^\alpha} {g_{s}(\gdot) - g_{0}}\right)}^{n^\prime} {\gdot}
 \label{eq:ss_strength_evolution}
\end{equation}
Here, a saturation strength appears that is assumed to depend on a net local plastic strain rate 
computed from the sum of the magnitudes of the slip system shearing rate:
\begin{equation}
g_{s}(\gdot) = g_1 \left( \frac{\gdot}{\gdots} \right)^{m^\prime} 
\hspace{0.5 cm} {\rm and} \hspace{0.5cm}
 \gdot = \sumss \arrowvert \gammadot \arrowvert 
 \label{eq:sat_ss_strength}
 \end{equation} 
This equation is used to update the strength of the slip systems, family-by-family, in each element of the 
finite element mesh used to discretize a polycrystal.

\subsection{Boundary conditions}
Consistent with solid mechanics theoretical framework, the boundary condition applied to a
surface of a virtual polycrystal may be either an imposed velocity or an imposed traction.  
For tractions, this is stated simply as:
\begin{equation}
\vctr{t}(\vctr{x})  =  \bar{\vctr{t} }
\label{eq:surface_traction}
\end{equation}
while for the velocity condition, it is:
\begin{equation}
\vctr{v}(\vctr{x}) =   \bar{\vctr{v}}
\label{eq:surface_velocity}
\end{equation}
where the overbar indicates a known quantity.  
\pagebreak[4]
\section{Finite Element Implementation}
\label{chapter:fem_implementation}
\subsection{Matrix notation for tensorial quantities}
To facilitate the presentation and implementation of the finite element formulation, tensor quantities are written as matrices.
Vectors map directly to one-dimensional column or row matrices.  For second order tensors, column vectors 
are defined for with a designated ordering to the components.  For the Kirchhoff stress and elastic strain, which are symmetric tensors, we use:
\begin{equation}
\kirch \rightarrow \left\{ \tau \right\} = \left\{ \tau_{11}\,\,  \tau_{22}\,\, \tau_{33}\,\, \sqrt{2}\tau_{23}\,\, \sqrt{2}\tau_{31}\,\, \sqrt{2}\tau_{12} \right\}^T
\label{eq:vector_stress}
\end{equation}
\begin{equation}
\lateps \rightarrow \left\{ \sf{e}^e \right\} = \left\{ \sf{e}^e _{11}\,\,  \sf{e}^e _{22}\,\, \sf{e}^e _{33}\,\, \sqrt{2}\sf{e}^e _{23}\,\, \sqrt{2}\sf{e}^e _{31}\,\, \sqrt{2}\sf{e}^e _{12} \right\}^T
\label{eq:vector_strain}
\end{equation}
where the $\sqrt{2}$ factor appears for the shear components in both tensors, which preserves the inner product
relation
\begin{equation}
\kirch  \cdot  \lateps = \left\{ \tau \right\}^T \left\{ \sf{e}^e  \right\} 
\label{eq:inner_product}
\end{equation}
For the deviatoric parts of the second order  tensors, a five-component form is adopted.  For the Kirchhoff stress and
the deformation rate:
\begin{equation}
\dkirch \rightarrow \left\{ \tau^\prime \right\} = \left\{ \frac{1}{\sqrt{2}}(\tau^\prime_{11}- \tau^\prime_{22})\,\,  \sqrt{\frac{3}{2}}\tau^\prime_{33}\,\,  \sqrt{2}\tau^\prime_{23}\,\, \sqrt{2}\tau^\prime_{31}\,\, \sqrt{2}\tau^\prime_{12} \right\}^T
\label{eq:vector_dev-stress}
\end{equation}
\begin{equation}
\dxdefrate \rightarrow \left\{ \sf{d}^\prime \right\} = \left\{ \frac{1}{\sqrt{2}}(\sf{d}^\prime_{11}- \sf{d}^\prime_{22})\,\,  \sqrt{\frac{3}{2}}\sf{d}^\prime_{33}\,\,  \sqrt{2}\sf{d}^\prime_{23}\,\, \sqrt{2}\sf{d}^\prime_{31}\,\, \sqrt{2}\sf{d}^\prime_{12} \right\}^T
\label{eq:vector_dev-defrate}
\end{equation}
\begin{equation}
{\tnsr{e}^e}^\prime \rightarrow \matdlateps = \left\{ \frac{1}{\sqrt{2}}(\sf{e}^\prime_{11}- \sf{e}^\prime_{22})\,\,  \sqrt{\frac{3}{2}}\sf{e}^\prime_{33}\,\,  \sqrt{2}\sf{e}^\prime_{23}\,\, \sqrt{2}\sf{e}^\prime_{31}\,\, \sqrt{2}\sf{e}^\prime_{12} \right\}^T
\label{eq:vector_dev-lateps}
\end{equation}
where the inner product again is preserved:
\begin{equation}
\dkirch  \cdot  \dxdefrate = \left\{ \tau^\prime \right\}^T \left\{ \sf{d}^\prime  \right\} 
\label{eq:inner_product_dev}
\end{equation}
Fourth-order tensors, namely the crystal elastic stiffness and compliance tensors, are commonly written as 6x6 matrices and populated according to the crystal symmetries.  
Hooke's law written using the matrix form in a crystal coordinate bases for cubic and hexagonal crystal types are:
\begin{equation}
\left\{
\begin{array}{c} \tau_{11} \\ \tau_{22} \\ \tau_{33} \\ \tau_{23} \\ \tau_{13} \\ \tau_{12} \end{array}
\right\} 
= 
\left[ 
\begin{array}{c c c c c c}
C_{11} & C_{12} & C_{12} &  &   &   \\ 
C_{12} & C_{11} & C_{12} &   &   &   \\ 
C_{12} & C_{12} & C_{11} &   &   &   \\ 
  &   &   & C_{44} &   &  \\  
  &   &   &  &  C_{44} &  \\ 
  &   &   &  &   &  C_{44}
\end{array}
\right] 
\left\{
\begin{array}{c} e_{11} \\ e_{22} \\ e_{33} \\ 2e_{23} \\ 2e_{13} \\ 2e_{12} \end{array}
\right\}   \hspace{0.5cm}{\rm Cubic \, Symmetry}
\label{eq:cubic_stiffness}
\end{equation}
and
\begin{equation}
\left\{
\begin{array}{c} \tau_{11} \\ \tau_{22} \\ \tau_{33} \\ \tau_{23} \\ \tau_{13} \\ \tau_{12} \end{array}
\right\} 
= 
\left[  
\begin{array}{c c c c c c}
C_{11} & C_{12} & C_{13} &  &   &   \\ 
C_{12} & C_{11} & C_{13} &   &   &   \\ 
C_{13} & C_{13} & C_{33} &   &   &   \\ 
  &   &   & C_{44} &   &  \\  
  &   &   &  &  C_{44} &  \\ 
  &   &   &  &   &  (C_{11}-C_{12})/2
\end{array}
\right] 
\left\{
\begin{array}{c} e_{11} \\ e_{22} \\ e_{33} \\ 2e_{23} \\ 2e_{13} \\ 2e_{12} \end{array}
\right\}   \hspace{0.5cm}{\rm Hexagonal \, Symmetry}
\label{eq:hexagonal_stiffness}
\end{equation}
where only the nonzero values are displayed.
   A cautionary remark is added here regarding a factor of 2 that may appear  with $C_{44}$ in other expressions of Hooke's law.   \fepx\, expects values for the elastic moduli consistent with Equation~\ref{eq:cubic_stiffness} or \ref{eq:hexagonal_stiffness} even though stiffness or compliance matrices internal to \fepx\, are constructed somewhat differently.   A restriction is placed on the moduli for hexagonal crystals vis-\`{a}-vis coupling of the shear and volumetric responses, as described in the next paragraph.

In the kinematic development, the motion is split into volumetric and deviatoric parts according to Equations~\ref{eq:kinematic_decomp_meandefrate} and \ref{eq:kinematic_decomp_devdefrate}.    This split is convenient for the
numerical implementation, but limits the generality of the hexagonal behaviors.   In particular, the volumetric and
deviatoric responses separate only when $C_{11}+C_{12} = C_{13}+C_{33}$.  Thus, only four of the five nonzero moduli
are independent.   To guarantee that this constraint is imposed, $(C_{11}, C_{12}, C_{13}, C_{44})$ are read in the input for \fepx\,  and $C_{33}$ is computed to satisfy the constraint.  

When split into volumetric and deviatoric parts, Equation~\ref{eq:hookes_law} gives:
\begin{equation}
\trace\matkirch =  \frac{\kappa}{3} \trace\matlateps 
\label{eq:volumetric_hooke}
\end{equation}
and
\begin{equation}
\matdkirch = \matxdelasticity\matdlateps
\label{eq:deviatoric_hooke}
\end{equation}
where the  stress and strain vectors are consistent with Equations~\ref{eq:vector_stress} - \ref{eq:vector_dev-defrate}.  
Table~\ref{tab:elasticmoduli} lists the values of the $\kappa$ and the nonzero (diagonal) entries of $\matxdelasticity$ in terms of the 
moduli presented in Equations~\ref{eq:cubic_stiffness} and \ref{eq:hexagonal_stiffness}.
\begin{table}[h]
\centering
\begin{tabular}{||c|c|c||}
\hline
\hline
Parameter & Cubic &Hexagonal  \\
\hline
$\kappa$  & $3(C_{11}+ 2C_{12})$  & $3(C_{11}+ C_{12}+C_{13})$  \\
$c^\prime_{11}$ & $C_{11}-C_{12}$  & $C_{11}-C_{12}$ \\
$c^\prime_{22}$   & $C_{11}-C_{12}$ & $3(C_{33}-C_{13})$  \\
$c^\prime_{33}$   & $C_{44}$ & $C_{44}$  \\
$c^\prime_{44}$   & $C_{44}$  & $C_{44}$ \\
$c^\prime_{55}$   & $C_{44}$ & $C_{11}-C_{12}$   \\
\hline
\hline
\end{tabular}
\caption{Values of the moduli used in the separated form of Hooke's law}
\label{tab:elasticmoduli}
\end{table}

\subsection{Time-discretized elastoplastic relations}
 Equations~\ref{eq:kinematic_decomp_meandefrate}, \ref{eq:kinematic_decomp_devdefrate},  \ref{eq:hookes_law} 
 and \ref{eq:ss_const_law} are now merged into a single equation that relates the Cauchy stress to the
 total deformation rate.   First,  the spatial time-rate change of the elastic strain is approximated with a finite difference expression:
\begin{equation}
\matlatepsdot = \deeteei \biggl( \matlateps - \matlatepsold \biggr)
\label{eq:euler_approx_strain_rate}
\end{equation}
where $\matlateps$ is the elastic strain at the end of the time step and $\matlatepsold$ is the elastic strain at the beginning of the time step.   The difference approximation is employed in an implicit algorithm, wherein the equations are solved at the time corresponding to the end of the time step.  This time corresponds to the current configuration.  
Writing the time rate change of the strain in terms of strains at two times facilitates substitution of Hooke's law -- namely at the end of the time step.   The elastic strain at the beginning of the time step is known from the solution for the preceding time step.
For the volumetric part of the motion, combining Equations~\ref{eq:kinematic_decomp_meandefrate} and \ref{eq:hookes_law}
with the difference expression gives:
\begin{equation}
-\pi  =  \frac{\kappa\Delta t }{ \beta } \trace\matdefrate +  \frac{\kappa} {\beta} \trace\matlatepsold
\label{eq:discret_volumetric_ep-law}
\end{equation}
Turning to the deviatoric (shearing) part of the motion,  inserting Equation~\ref{eq:euler_approx_strain_rate} into Equation~\ref{eq:kinematic_decomp_devdefrate} gives:
\begin{equation}
\matddefrate = \deeteei \matdlateps + \matlatdefrate + \matpxspinhat \matdlateps - \deeteei \matdlatepsold
\label{eq:discret_deviatoric_ep-law}
\end{equation}
where $\matpxspinhat$ is the matrix form of $\pxspinhat$:
\begin{equation}
\matpxspinhat = 
\left[
\begin{array}{c c c c c}
0 &  0 & -2{\hat w_{12}^p} & -{\hat w_{13}^p} &  {\hat w_{23}^p} \\ 
0 & 0 & 0 & {\sqrt{3}} {\hat w_{13}^p} & {\sqrt{3}} {\hat w_{23}^p}  \\ 
2{\hat w_{12}^p}& 0 & 0 & -{\hat w_{23}^p}  &  -{\hat w_{13}^p}  \\ 
{\hat w_{13}^p}  & -{\sqrt{3}}{\hat w_{13}^p}  &  {\hat w_{23}^p} & 0  &  -{\hat W_{12}^p} \\  
-{\hat w_{23}^p}  & -{\sqrt{3}}{\hat w_{23}^p}  &  {\hat w_{13}^p} & {\hat w_{12}^p}   &  0 
\end{array}
\right] 
\end{equation}
The equations for plastic slip 
(Equation~\ref{eq:ss_const_law}) for the plastic deformation rate:
\begin{equation}
\matlatdefrate = \matxplasticity \matdkirch
\end{equation}
\begin{equation}
\matxplasticity
= 
\sumss \left( \frac{f(\rss, g)}{\rss} \right) 
\matsymschmid  \matsymschmid^\transp
\end{equation}
together with Equation~\ref{eq:deviatoric_hooke} 
are substituted to render an equation that gives the deviatoric Cauchy stress in terms of the total deviatoric deformation rate and 
a matrix, $\mathhh$, that accounts for the spin and the elastic strain at the beginning of the time step: 
\begin{equation}
\matdcauchy  =  \matxep \bigg( \matddefrate- \mathhh \bigg)
\label{eq:discrete_devCauchy}
\end{equation}
where:
\begin{equation}
\matxep^{\invrs} = \frac{\beta}{\Delta t} {\matxdelasticity}^{\invrs} +\beta \matxplasticity 
\label{eq:discrete_ep_stiffness}
\end{equation}
\begin{equation}
\mathhh = \matpxspinhat  \matdlateps - \deeteei \matdlatepsold
\label{eq:discrete_spin_correction}
\end{equation}
Equations~\ref{eq:discrete_devCauchy}, \ref{eq:discrete_ep_stiffness} and \ref{eq:discrete_spin_correction} will be used in the weak form of equilibrium to write the stress in terms of the deformation rate.

\subsection{Interpolation functions}
\fepx employs a standard isoparametric mapping framework for discretizing the problem domain and for representing the solution variables.  
The mapping of the coordinates of points provided by the elemental interpolation functions, $\matcapN$, and the coordinates of the nodal points, $\left\{ X \right\}$:
\begin{equation}
\left\{ x \right\}  = \matcapN  \left\{ X \right\}
\label{eq:coord_mapping}
\end{equation}
where $(\xi, \eta, \zeta)$ are local coordinates within an element.  
The same mapping functions are used for the solution (trial) functions which, together with the nodal point values of the
velocity, $\matvelnp$, specify the velocity field over the elemental domains:
\begin{equation}
\matvel  = \matcapN  \matvelnp
\label{eq:trial_functions}
\end{equation}
The deformation rate is computed from the spatial derivatives (derivatives with respect to $\vctr{x}$) of the mapping functions and the nodal velocities as:
\begin{equation}
\matdefrate = \matcapB \matvelnp
\label{eq:trial_function_derivatives}
\end{equation}
$\matcapB$ is computed using the derivatives of $\matcapN$ with respect to local coordinates, $(\xi, \eta, \zeta)$, together with the Jacobian of the mapping specified by Equation~\ref{eq:coord_mapping}, following standard finite element procedures for isoparametric elements.

\begin{figure}[h]
\begin{center}
\includegraphics*[width=10cm]{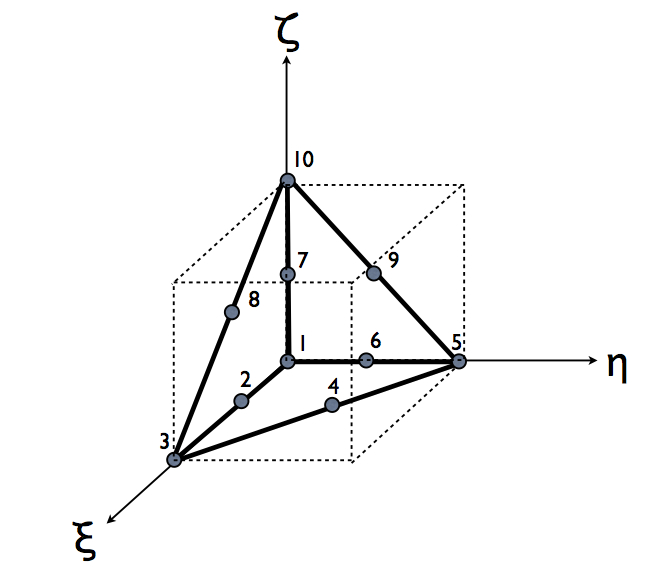}
\caption{10-node tetrahedral element with quadratic interpolation of the velocity, shown in the parent configuration and bounded by a unit cube.}
\label{fig:tet_element}
\end{center}
\end{figure}

\fepx\, relies principally on a 10-node, tetrahedral, serendipity element, as shown in Figure~\ref{fig:tet_element}.  This $C^0$ element provides pure  quadratic interpolation of the velocity field.  
\fepx\, employs a Galerkin methodology for constructing a weighted residual.  The weight functions therefore use the same interpolation functions as used for the coordinate map and the trial functions:
\begin{equation}
\Big\{ \psi \Big\} =  \matcapN  \Big\{ {\it \Psi } \Big\}
\label{eq:weight_functions}
\end{equation}

\subsection{Finite element residual for the velocity field}
\label{sec:fe_formulation}
Equilibrium is enforced by requiring a global weighted residual to vanish:
\begin{equation}
R_u = \int_{\cal{B}} \gtnsr{\psi} \cdot \left( \divop {\cauchy}^T + {\vctr{\iota}}\right) \dee {\cal B} = 0
\end{equation}
The residual is manipulated in the customary manner (integration by parts and application of the divergence theorem) to obtain the weak form:
\begin{equation}
R_u = 
- \int_{\cal{B}} \; \trace  \left( {\dcauchy}^\transp  \,\gradop \vctr{\psi} \right) \dee {\cal{B}} 
+  \int_{\cal B} \pi \, \divop \vctr{\psi} \dee {\cal B}
+ \int_{\partial {\cal B}} \vctr{t} \cdot \vctr{\psi} \dee\Gamma +
\int_{\cal{B}} \vctr{\iota} \cdot \vctr{\psi} \dee{\cal B}
\end{equation}
Introduction of the trial and weight functions gives a  residual vector for the discretized weak form for each element:  
\begin{equation}
\matresiduale
=
\bigg[ \matstiffd + \matstiffv \bigg] \matvelnp
-
\Big\{ \mathsf{f}^{\it ele}_a \Big\}
-
\Big\{ \mathsf{f}^{\it ele}_d \Big\}
-
\Big\{ \mathsf{f}^{\it ele}_v \Big\}
\end{equation}
where
\begin{equation}
\matstiffd 
= 
\int_{\cal{B}} \matcapB^\transp  \matcapX^\transp \matxep \matcapX \matcapB \dee {\cal B}
\label{eq:k_d}
\end{equation}
\begin{equation}
\matstiffv = 
\int_{\cal{B}} \betaoverkappa \matcapB^\transp \matcapX^\transp \matdelta \matdelta^\transp \matcapX \matcapB \dee {\cal B}
\end{equation}
\begin{equation}
\Big\{ \mathsf{f}^{\it ele}_a \Big\} 
= 
\int_{\partial {\cal{B}}} \matcapN^\transp \mattraction \dee \Gamma  
+ \int_{\cal{B}} \matcapN^\transp \matbodyforce \dee {\cal B}
\end{equation}
\begin{equation}
\Big\{ \mathsf{f}^{\it ele}_v \Big\} 
= \int_{\cal B} \matcapB^\transp \matcapX^\transp \betaoverkappa \matdelta^\transp \matlatepsold \dee {\cal B} 
\end{equation}
\begin{equation}
\Big\{ \mathsf{f}^{\it ele}_d \Big\} 
=  \int_{\cal B} \matcapB^\transp \matcapX^\transp \matxep \mathhh \dee {\cal B} 
\label{eq:f_d}
\end{equation}
The integrals appearing in Equations~\ref{eq:k_d}-\ref{eq:f_d}  are evaluated by numerical quadrature.  
%\parbox[t]{10cm}{
%\begin{center}
% {\bf  ADD:  Insert discussion of quadrature rules.  Add table with quad points and weights.  }
%\end{center}
%}

Assembling the elemental matrices and requiring that the residual vanish for all independent variations in the 
weights gives:
\begin{equation}
\bigg[ \Matstiffd + \Matstiffv \bigg] \matvelnp
=
\Big\{ \mathsf{F}_a \Big\}
+
\Big\{ \mathsf{F}_d \Big\}
+
\Big\{ \mathsf{F}_v \Big\}
\label{eq:global_system}
\end{equation}
The essential boundary conditions are applied prior to solving for the nodal velocities, as described in Section~\ref{sec:fe_boundary-conditions}.  The matrix equation given by Equation~\ref{eq:global_system} is nonlinear ($\Matstiffd$ and $\Matstiffv$ depend on $\matvelnp$).  The solution methodology used in \fepx\, is outlined in Section~\ref{sec:fe_nonlinear_methods}.

\subsection{Nonlinear solution algorithm for obtaining the velocity field}
\label{sec:fe_nonlinear_methods}
To solve Equation~\ref{eq:global_system} for the velocity field, an iterative methodology is invoked.   This methodology is a hybrid procedure that utilizes a combination of successive approximations (Picard) and Newton-Raphson updates.   
To accomplish this, the assembled residual force vector, $\matresidual$, is defined as:
\begin{equation}
\matresidual =  \bigg[ \Matstiffd + \Matstiffv \bigg] \matvelnp
-
\Big\{ \mathsf{F}_a \Big\}
-
\Big\{ \mathsf{F}_d \Big\}
-
\Big\{ \mathsf{F}_v \Big\}
\label{eq:global_residual}
\end{equation}
The goal of the iterative process is to drive the residual to zero through a series of corrections, $\delmatvelnp$, to an estimate of the velocity field.   Denoting the estimate of the velocity on the $i^{\rm th}$ iteration as $\matvelnp^{i}$ and the estimate on the next iteration as $ \matvelnp^{i+1}$, the iteration procedure is written simply as: 
\begin{equation}
 \matvelnp^{i+1}   =   \matvelnp^{i} +  \delmatvelnp^{i+1}
\label{eq:iterative_velocity}
\end{equation}
where $\delmatvelnp^{i+1}$ is determined from the solution of
\begin{equation}
\bigg[ \Matstiffd^{{\rm type}} + \Matstiffv \bigg]  \delmatvelnp^{i+1}   =  -{\matresidual}^{i}
\label{eq:velocity_iterations}
\end{equation}
Here, $\Matstiffd^{{\rm type}}$ refers to either a tangent modulus, $\Matstiffd^{{\rm tan}}$ or a secant modulus, $\Matstiffd^{{\rm sec}}$, as specified by the hybrid procedure.   
Convergence is based on changes in the norm of $\matvelnp$ becoming small.  

\subsection{Time marching and boundary conditions}
\label{sec:fe_boundary-conditions}

The intent of the \fepx\, framework and the derivative finite element code is to simulate the deformation of virtual polycrystals over time.  To this end, time histories are approximated 
by solving for the velocity field at a series of discrete times.     Simply stated, \fepx\, computes the velocity field at the end of time interval using Equation~\ref{eq:global_residual} knowing the velocity field and state at beginning of the time interval.    
The geometry and state variables (lattice orientations and slip system strengths) are updated concurrently with
 the velocity field at the end of the time step.
The time marching method is documented in ~\cite{mar_daw_98b,man_daw_lee_92}.

Over the course of a deformation history, the applied boundary conditions often change.  This may imply that either
natural or essential boundary conditions in Equations~\ref{eq:surface_traction} and \ref{eq:surface_velocity} are functions of time.  Presently, \fepx\, allows the user to change the values of imposed velocities or forces at nodes over the course of a deformation, but does not allow the user to change the type of boundary condition.   That is, a nodal point with imposed velocity (essential boundary condition) will have an essential boundary condition throughout a simulation, but the value of the velocity that is imposed may change with time.   The same applies for nodes with natural boundary conditions  -- the force may change with time, but the condition at that node will remain a natural boundary condition  throughout the simulation.  

Given this limitation,   the boundary condition options within \fepx\, accommodate a number of possibilities.   For example,
it is anticipated that a common use of \fepx\, will be to simulate the response of virtual polycrystals being subjected to boundary conditions that replicate mechanical tests performed using a load frame.  
The control of mechanical tests can be designed to provide
programmed force histories,  programmed displacement histories, or combinations of these.  \fepx\, offers the
capabilities to impose boundary conditions to mimic mechanical test histories.  Some of the possibilities include:
\begin{itemize}
\item  specified crosshead/actuator velocity,
\item  specified load history,
\item  unloading episodes, and
\item  uniaxial and  biaxial loading modes.
\end{itemize}

\pagebreak[4]
\section{Input and Output Data}
\label{chapter:IO}
The modeling framework described in this article is intended for 
simulating the deformations of polycrystalline aggregates.  
The material state (including phase topology, grain morphology, crystallographic texture, slip system strengths) both influences the deformation and is affected by the deformation, and thus is 
an integral part of the definition of a material system being modeled.  The finite element formulation
provides the solution methodology for solving the system of equations under imposed 
loading for the polycrystalline aggregate of interest.   
 Thus, the input and output data for \fepx\, are similar in content to other finite element codes, but 
are tailored for and organized around the polycrystals.   
One must define the finite element mesh, assign material attributes consistent with the state to the elements, and specify boundary and initial conditions.  
In addition, information needed for internal computational procedures, such as choices regarding type of nonlinear solver to employ,
convergence tolerances on nonlinear iterations, as well as the frequency and extent of output data must be provided.
A general description of the input data is given here.  The detailed description of the input data needed by a user
to execute \fepx\, is given in a separate document.    

Because \fepx\, uses microstructural information generated by experiment or simulation to assign material attributes, 
pre-processing routines are needed to prepare input files.  Other software package serve this function.   \odfpf\, is a suite of 
Matlab routines written to perform a variety of tasks related crystallographic texture and is recommended as a complement to \fepx.
\neper\, is capable of meshing  virtual polycrystals created by Voronoi tesselation and constructs the nodal point and element data needed by \fepx.  Instructions for using these packages are included with the respective packages.    Other packages for visualization,   interpretation and archiving data also are needed for post-processing \fepx\, results.   Again, \odfpf\, provides capabilities to translate \fepx\, output files into formats appropriate for using other software packages.

\subsection{\fepx\, input data }\label{sec:input_data}
The body being loaded and deformed in a \fepx\, simulation is a virtual polycrystal  -- a set of grains (each grain being a single crystal) that forms a fully-dense solid.    The finite element mesh needed to build a virtual polycrystal must be created in advance, which can be done with \neper\, or other mesh generation codes.
The input data is organized as follows:
\begin{itemize}
\item {\bf Setting up a job --} a script to execute \fepx\, together with a file which lists the input files described below. 
\item {\bf Defining a virtual polycrystal --}  a set of files: a file that specifies the single crystal elastic and plastic material properties for each phase; a file that defines the finite element mesh (nodal coordinates, element connectivity, and surface elements); a file that designates  the phase and grain numbers for each element; a file that provides the starting lattice orientation for each grain; and, one or more files that define the vertices of the single crystal yield surface for each phase.  
\item  {\bf Controlling the deformation --} a set of two files: a file designates the type of boundary condition (essential or natural) for each degree of freedom of all the nodes and the corresponding velocity or force; and, a file specifies target loads or displacements, depending on the loading type.  
\item  {\bf Postprocessing for diffraction data --} a file that provides  information for averaging strains and stresses over crystallographic fibers. 
\item  {\bf Prescribing optional input --} a file that specifies the choices for loading protocols, post-processing options, various solution method options,  and associated convergence limits. 
 
\end{itemize}
   
\subsection{\fepx\, output data}
\label{sec:output_data}   
In the present implementation, the \fepx\, code writes files for solution variables at times designated in the input data.  The solution variables written to files are:
\begin{itemize}
\item  {\bf Geometry:}   the nodal point coordinates and the nodal point velocities.  
\item  {\bf Deformation:}  the effective deformation rate, the effective plastic deformation rate, the slip system shearing rates,  the effective strain, and the effective plastic strain.
\item  {\bf State Variables:}   the slip system strengths and the lattice orientations. 
\item  {\bf Stress:}  the stress tensor.    Note that the components of the Cauchy stress tensor are written in the global frame 
in the following order: $ \sigma_{11},    \sigma_{12} ,   \sigma_{13}, \sigma_{22},     \sigma_{23}, \sigma_{33}   $.
\item  {\bf Elastic strain:}  the elastic strain tensor.    As with the stress, the components of the elastic strain tensor are written in the global frame 
in the following order: $ \epsilon_{11},   \epsilon_{12},   \epsilon_{13},    \epsilon_{22},   \epsilon_{23},  \epsilon_{33}$.
\item {\bf Fiber averages:} mean and standard deviation values of the stress, elastic strain, and slip system activities  taken over designated crystallographic fibers.
\item {\bf Monitors:}   the resultant force on each external surface.  
\end{itemize}
When executed on a parallel architecture, \fepx\, writes output files on every node which are written to the master node at the end of the job.   This leaves the results spread over many output files and in a format that is inconvenient for postprocessing.  An \odfpf\, script is available to concatenate all of the output files into a single data structure that is readily used within postprocessing tools.  
  
\subsection{Exporting  Input and Output Data }
\label{sec:hdf5_io}   
Exporting input and output data is done for two reasons.   One is so that the results may be archived in a data management system appropriate for the project of interest.   The other is to interface with visualization codes or other specialty post-processing software, such as forward projectors or virtual instruments.    The \odfpf\, package includes scripts for these purposes:  
\begin{itemize}
\item   The HDF-5 file format is a commonly used standard for writing archivable files.  A matlab script is available that prepares an HDF-5 file with the \fepx\, standard input and output data.   This script can be modified to include other information (such as postprocessing results) if desired. 
\item  To facilitate visualization of the results, matlab scripts are available to export  files that can be read in by \paraview\, or \visit.  The applications of \fepx\, shown in Section~\ref{chapter:examples} were plotted with \paraview.
\end{itemize}
   
\subsection{\odfpf\, capabilities}
\label{sec:odfpf}
\odfpf\, figures prominently in the use of \fepx\, to simulate the behavior of virtual polycrystals, both in the instantiation of virtual polycrystals and in the comparison of simulation results to diffraction data.  \odfpf\,   is a function set is a collection of \matlab\, functions which operate on ODF's (orientation distribution functions) and PF's (pole figures). It handles plotting of the ODF using Rodrigues parameters, plotting of pole figures and inverse pole figures, evaluation of pole figures inverse pole figures from ODF's, and it provides many tools for computing ODF's from pole figures. 
Archival publications are available that cover various aspects of its use or the use of a Rodrigues parameterization of  orientation space in quantitative texture analyses.  Relevant articles include: \cite{kum_daw_00,bar_boy_daw_02, bernierodf_etal_2006,schurenUncert_2011}.

\pagebreak[4]
\section{Example Problems}
\label{chapter:examples}
To illustrate the use of \fepx\,  in simulating the mechanical response of virtual polycrystals, three example problems are presented.  
We use  different methods for instantiating  virtual polycrystals for each of the three examples.    The first is an example of building
a virtual polycrystal comprised of regular, dodecahedral-shaped grains.  In the second example, the virtual polycrystal is a discretized Voronoi tessellation.   The third example demonstrates the construction of a virtual polycrystal by mapping serial sectioning data directly onto a uniform mesh.     All examples were executed on a parallel computer using 64 processors (8 nodes each with 8 cores).  The same version of \fepx\, was used for all examples with data files conforming to the input manual.   Postprocess of the results was performed with \odfpf\, scripts to write  *.vtk files.  
The *.vt files were imported into  \paraview\, for plotting. 

\subsection{Virtual Polycrystal Generated by Regular Tessellation }
\label{sec:dodecahedramesh}
This example was provided by Andrew Poshadel  and was developed as part of his research on yielding of dual phase alloys~\cite{poshadel_phd}. 
The example demonstrates the application of \fepx\, to a virtual polycrystal with two phases that was built using 
dodecahedral grains.  The two phases are the austenitic and ferritic phases of a dual steel (LDX-2101), referred to symbolically as the $\gamma$ and $\alpha$ phases, respectively.   The stock material has a microstructure consistent with having been rolled or extruded.

\subsubsection{Defining the virtual polycrystal}
Following the summary of the input data given in Section~\ref{sec:input_data}, the required input data to execute \fepx\, can be organized into five categories:
\begin{enumerate}
\item{{\bf Phase Attributes:}   The  $\alpha$ and  $\beta$ phases both have cubic crystal structure  -- FCC for the $\gamma$ phase and BCC for the $\alpha$ phase.  The single crystal, cubic, elastic moduli for the two phases are listed in Table~\ref{tab:example1_elasticmoduli}.   Based on experimental data that indicates the plastic behaviors of the two phases are comparable,   the same plasticity parameters were assigned to both phases for the purpose of this example.  These parameters are listed in Table~\ref{tab:example1_plasticparms}.    The slip systems are different for the two, however, with the FCC $\gamma$ phase using the $\{111\} \, <110>$ systems 
and the BCC $\alpha$  phase using the $ \{110\} \, <111>$ systems  (See Table~\ref{tab:slip_systems}).
This information is provided to \fepx\, in the *.matl input file.}
\begin{table}[h]
\centering
\begin{tabular}{||c|c|c|c|c||}
\hline
\hline
Phase &  Type & $C_{11}$  (GPa)  & $C_{12}$  (GPa)  & $C_{44}$  (GPa)  \\
\hline
$\gamma$ & FCC   & 204.6 & 137.7 & 126.2 \\
$\alpha$  & BCC    & 236.9 & 140.3 & 116.0 \\
\hline
\hline
\end{tabular}
\caption{LDX 2101 elastic moduli used in the single crystal constitutive equations for the two-phase virtual polycrystal.  
Values listed conform to the convention given in Equation~\ref{eq:cubic_stiffness}. }
\label{tab:example1_elasticmoduli}
\end{table}
 \begin{table}[h]
\centering
\begin{tabular}{||c|c|c|c|c|c|c|c|c||}
\hline
\hline
Phase & $\gdotz$ (${{\rm s}^{-1}}$) & $m$ & $h_{0}$ (MPa) & $g_0$ (MPa)& $n^\prime$  & $g_1$ (MPa)  & $\gdots$ (${{\rm s}^{-1}}$) & $m^\prime$\\
\hline
$\gamma$ &  1.0 & 0.02 & 391.9 & 237.0 & 1 & 335.0  & $5.0\times10^{10}$ & 0  \\
$\alpha$  &   1.0 & 0.02 & 391.9 & 237.0 & 1 & 335.0  & $5.0\times10^{10}$ & 0  \\
\hline
\hline
\end{tabular}
\caption{LDX 2101 slip parameters used in the single crystal constitutive equations for the two-phase virtual polycrystal.
Values listed conform to the convention given in Equations~\ref{eq:ss_kinetics}, \ref{eq:ss_strength_evolution}, and
\ref{eq:sat_ss_strength}.} 
\label{tab:example1_plasticparms}
\end{table}
\item{{\bf Mesh definition:}  A finite element mesh underlying the virtual polycrystal was instantiated using a custom \matlab\, script.  It consists of a regular, rectangular layout of elements.   These elements can be grouped to form regular dodecahedral grains.    The resulting mesh, shown in Figure~\ref{fig:example1_mesh}, has 117,504 10-node tetrahedral elements and 173,829 nodal points.
The corresponding arrays for the nodal point coordinates and element connectivities are  provided to \fepx\, in the *.mesh input file.   }
 \begin{figure}[h!]
\centering

  \includegraphics*[width=10cm]{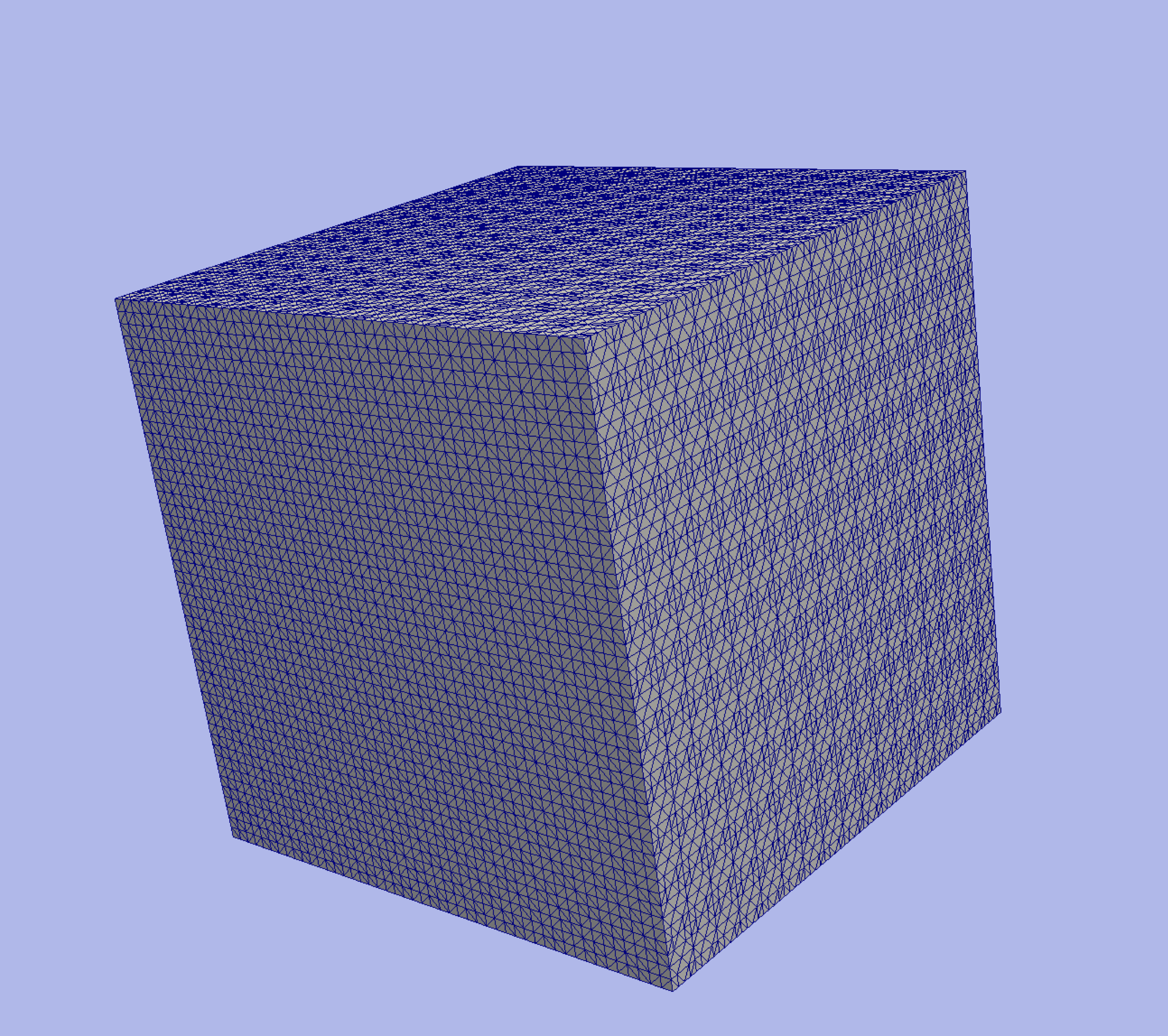} 

\caption{Finite element mesh for the dual phase steel virtual sample. }
\label{fig:example1_mesh}
\end{figure}

\item{{\bf Phase and grain definition:} The finite elements must be assigned to phases and grains.  The first step was to define the  spatial distributions of the two phases in one plane using a regular layout of dodecahedral grains.    
The second step was to extrude (repeat) the planar layout in the direction perpendicular to the plane to create the cube-shaped polycrystal with a microstructure similar to the stock material.    In this example, elements are one of the two phases ($\gamma$ or  $\alpha$).   Within subdomains of a single phase, there may be one or more crystals.    For this dual phase steel, the subdomains of both phases  have multiple crystals, although the $\gamma$ phase subdomains generally had fewer crystals than the subdomains of the  $\beta$ phase subdomains.   The grain definitions follow from the assignment of lattice orientations to the elements.   Contiguous elements with the same orientation constitute a grain.  Lattice orientations were chosen randomly from measured orientation distributions and assigned to element to create the desired grain arrangement within the phases.
%The orientation distributions for the two phases are shown in Figure~\ref{fig:example1_phase-and-grain}.
Each grain also was assigned the same initial slip system strength, which in turn was assigned to every element of the grain.  
Figure~\ref{fig:example1_phase-and-grain} shows the phase and grain assignments associated with the mesh shown in Figure~\ref{fig:example1_mesh}.
This information is provided in the *.grains and *.kocks files.    }
\begin{figure}[h!]
\centering
\subfigure[Phases]
{
  \includegraphics*[width=8cm]{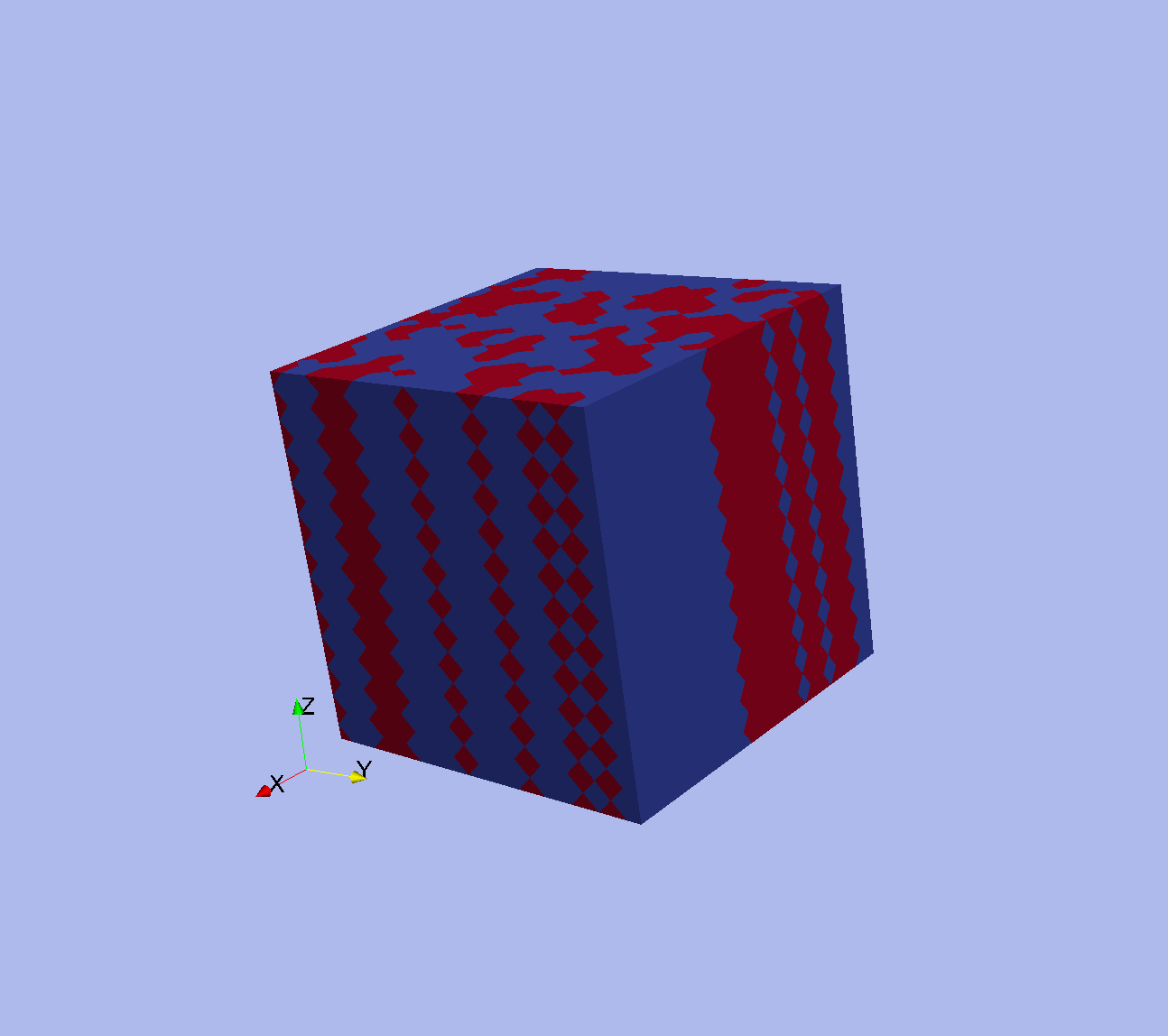} 
}
\subfigure[Grains]
{
   \includegraphics*[width=8cm]{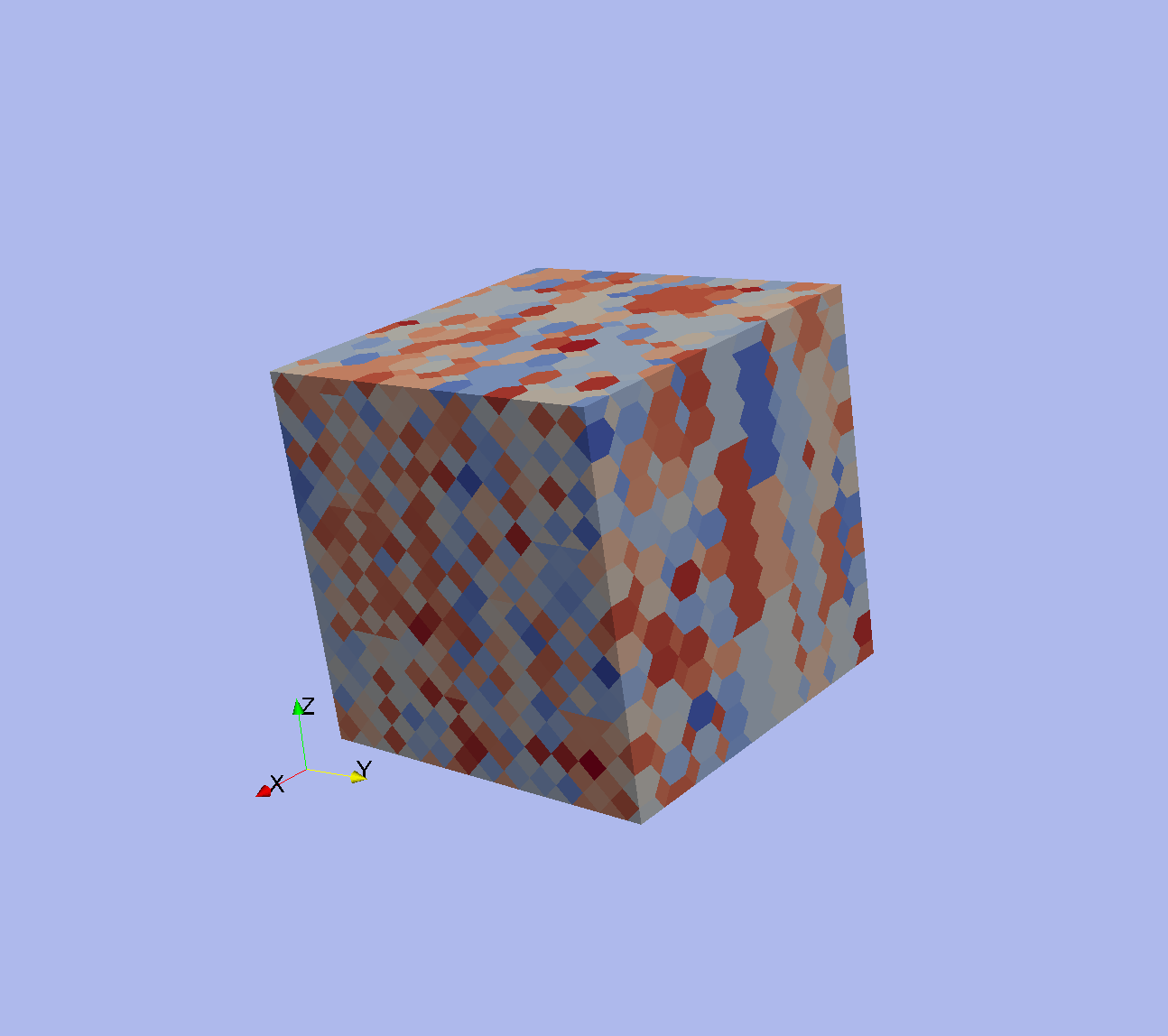}
}
\caption{Phase and grain distributions for the dual phase steel virtual sample.  For the phase distribution, blue indicates the BCC ferritic ($\alpha$) phase; red indicates the FCC austenitic ($\gamma$) phase. For the grain distribution, grains are indicated by domains of uniform color.}
\label{fig:example1_phase-and-grain}
\end{figure}
\item{{\bf Vertex files:} standard definitions of the single crystal yield surface vertices were used for both the FCC and BCC phases.} 
\end{enumerate}
\subsubsection{Controlling the loading:}
\begin{enumerate}
\item{{\bf Boundary conditions:}  The boundary conditions are intended to simulate the loading applied in a tension test.
The virtual polycrystal is constrained on the bottom and stretched in the z direction by an imposed axial velocity on the top.
Two adjoining lateral surfaces are traction-free  while a symmetry condition is applied  on the other two.    Rigid body translations and rotations have been suppressed by the application of the symmetry conditions.  This information is given in the *.bcs  file.  }
\item{{\bf Target loads:}  Simple load control is applied to extend the sample.   Several z-direction target loads along a monotonically increasing path (no unloading episodes) are specified to provide points for writing output data.    This information is given in the  *.loads file.}
\end{enumerate}

\subsubsection{Specifying options:}
\begin{enumerate}
\item{{\bf Load controls:}   The ``control by load'' mode is used to control the loading history. }
\item{{\bf Convergence criteria:} The default parameters have been used for both the velocity field and crystal stress iterative procedures.   The Newton-Raphson procedure is invoked for the velocity solutions.}
\end{enumerate}

\subsubsection{Selected Simulation Results}
\begin{figure}[h!]
\centering
\subfigure[Axial stress component. ]
{
  \includegraphics*[width=8cm]{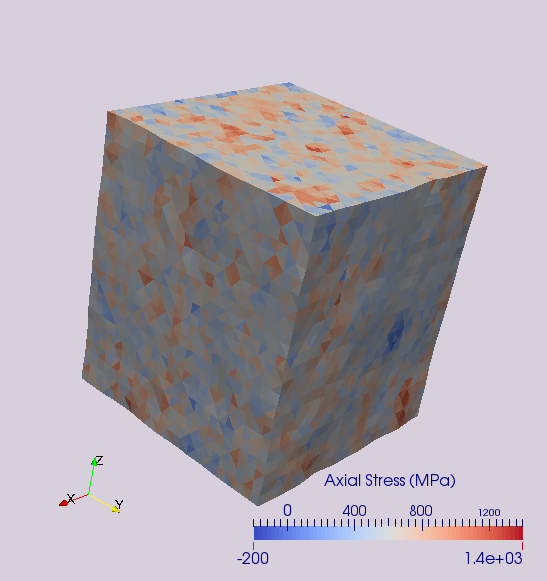} 
}
\subfigure[Effective plastic strain. ]
{
   \includegraphics*[width=8cm]{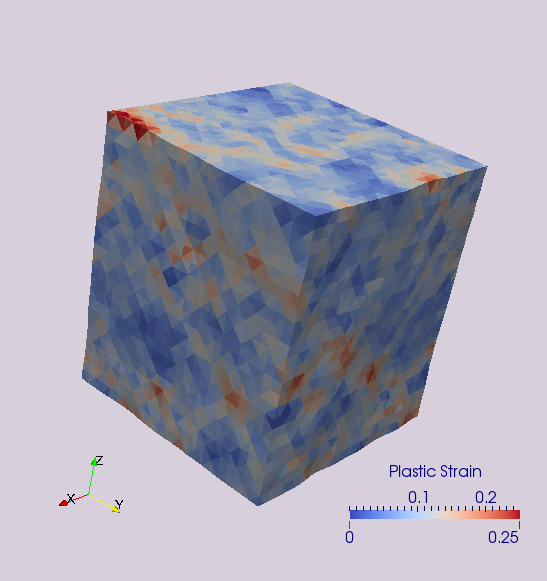}
}
\caption{Axial stress and plastic strain distributions  at nominal load of 590N shown on the deformed mesh.}
\label{fig:example1_stresses}
\end{figure}

\begin{figure}[h!]
\centering
  \includegraphics*[width=6cm]{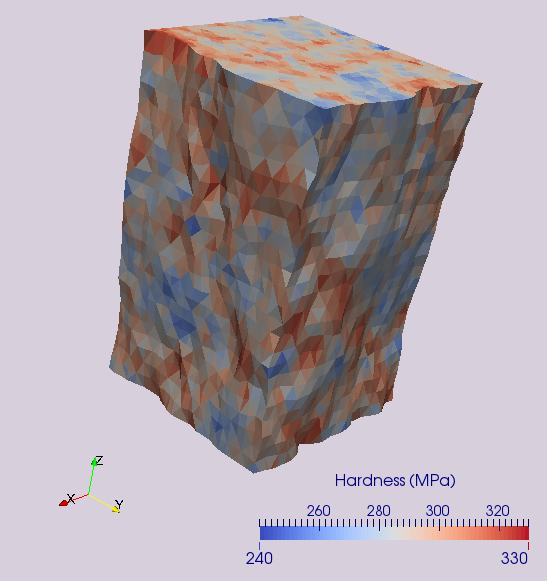} 
\caption{Slip system strength at 590N shown on a exaggerated (x4) deformation field.}
\label{fig:example1_hardness}
\end{figure}

Simulation results are available for postprocessing at the points in the loading designated by the target loads.   
These results were aggregated and written to a .vtk file for plotting with \paraview.  
Figure~\ref{fig:example1_stresses} shows the distribution of axial component of the stress and the effective plastic strain at the last target load of 590 N.
At this load, the nominal axial strains was approximately 10\%.
The stress shows spatial variations due to the anisotropy of the crystal properties and the 
interactions among the grains.     There is an  increase in the variability of the axial component of the stress as the stress level is increased due in part to the change in the principal directions of the crystal stresses as the stresses move toward a vertex of the single crystal yield surface during the elastic-plastic transition.   The effective stress (not show here) exhibits less variability.       
Figure~\ref{fig:example1_hardness} shows the evolution of strength over the course of the loading.  The deformation of the mesh has been exaggerated by a factor of 4  to facilitate visualizing the heterogeneity of the deformation.
A correlation between the plastic deformation and the strain hardening is evident.
A great deal more information is available in the simulation output.  For example, lattice strains in crystals on designated crystallographic
fibers was collected and averaged for comparisons to experiments in which lattice strains were measured by  neutron diffraction during
{\it in situ} loading.
%\pagebreak
   
\subsection{Voronoi Tessellated Virtual Polycrystal Generated with \neper}
\label{sec:voronoimesh}
This example was provided by Amanda Mitch and was developed as part of her research on reduced-order representation of
crystal stress distributions for use in a methodology for quantifying residual stress distributions in engineering components~\cite{mitch_ms}. 
The example demonstrates the application of \fepx\, to a virtual polycrystal  that was built using 
a Voronoi tessellation to define the grains.  The material is single phase and the single crystal properties are consistent with a FCC crystal type, but do not
represent any particular metal or alloy.  
The input data, following the organization given in Section~\ref{sec:input_data}, is summarized below.

\subsubsection{Defining the virtual polycrystal}
\begin{enumerate}
\item{{\bf Phase Attributes:} 
The material is single phase with a  cubic (FCC) crystal structure.  The single crystal, cubic, elastic moduli  are listed in Table~\ref{tab:example2_elasticmoduli}.    Plasticity parameters are generic, being similar to copper alloy.  These parameters are listed in Table~\ref{tab:example2_plasticparms}.    The slip systems are the customary primary systems for FCC crystals: the $\{111\} \, <110>$ systems.
This information is provided to \fepx\, in the *.matl input file.
  \begin{table}[h]
\centering
\begin{tabular}{||c|c|c|c|c||}
\hline
\hline
Phase &  Type & $C_{11}$  (GPa)  & $C_{12}$  (GPa)  & $C_{44}$  (GPa)  \\
\hline
$\alpha$& FCC                & 245. & 155. & 62.5 \\
\hline
\hline
\end{tabular}
\caption{Elastic moduli used in the single crystal constitutive equations for the Voronoi-based virtual polycrystal.}
\label{tab:example2_elasticmoduli}
\end{table} 
 \begin{table}[h]
\centering
\begin{tabular}{||c|c|c|c|c|c|c|c|c||}
\hline
\hline
Phase & $\gdotz$ (${{\rm s}^{-1}}$) & $m$ & $h_{0}$ (MPa) & $g_0$ (MPa)& $n^\prime$ & $g_1$ (MPa)   & $\gdots$ (${{\rm s}^{-1}}$) & $m^\prime$\\
\hline
$\alpha$ &  1.0 & 0.05 & 200. & 210.& 1  & 330.& $5.0\times10^{10}$ & $5.0\times10^{-3}$  \\
\hline
\hline
\end{tabular}
\caption{Slip parameters used in the single crystal constitutive equations for the Voronoi-based virtual polycrystal.}
\label{tab:example2_plasticparms}
\end{table}
}
\item{{\bf Mesh definition:} 
A virtual  polycrystal was instantiated using the \neper\, code.  \neper\, builds a Voronoi construction of the domain to define grains and then discretizes the grains into finite elements.   The resulting mesh, shown in Figure~\ref{fig:example2_mesh_grains}, has 96,758 10-node tetrahedral elements and 134,362  nodal points.
The corresponding arrays for the nodal point coordinates and element connectivities are  provided to \fepx\, in the *.mesh input file.   
 \begin{figure}[h!]
\centering
\subfigure[Mesh]
{
  \includegraphics*[width=6cm]{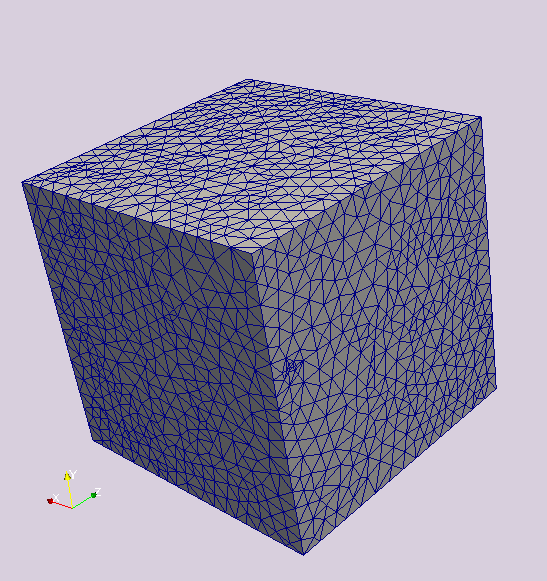}
}
\subfigure[Grains]
{
  \includegraphics*[width=6cm]{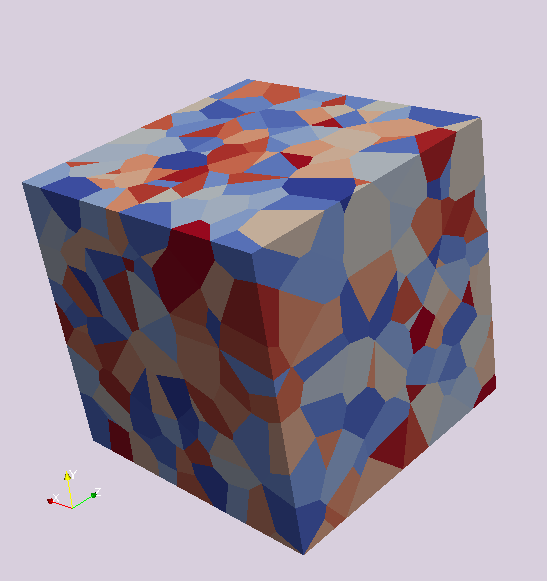}
}
\caption{Mesh and grains for the \neper-built polycrystal. }
\label{fig:example2_mesh_grains}
\end{figure}
}
\item{{\bf Phase and grain definition:}
All grains are the same phase.  The grain designations for the finite elements follow from
the Voronoi tessellation and are shown in Figure~\ref{fig:example2_mesh_grains}.  
This information is provided in the *.grain and *.kocks files. 
}

\item{{\bf Vertex files:} 
a standard definition for the vertices of a FCC single crystal yield surface was used.} 
\end{enumerate}

\subsubsection{Controlling the loading:}
\begin{enumerate}
\item{{\bf Boundary conditions:}  The boundary conditions are intended to simulate the triaxial loading of the sample such that the
stress components remain in constant proportions given by.:
\begin{equation}
[\sigma]
= 
\sigma_1 \left[ 
\begin{array}{c c c }
1.0 & 0 & 0  \\ 
0 & -0.625 & 0 \\ 
0 & 0 & -0.375  
\end{array}
\right] 
\label{eq:triaxial_stress_state}
\end{equation}

Normal velocity components of different magnitudes are applied to three adjacent surfaces while the opposing surfaces are fixed in place.
The surface velocities are adjusted to achieve tractions consistent with the target ratios for triaxial stress state.  }
\item{{\bf Target loads:}    Two target loads along a monotonically increasing path (no unloading episodes) are specified to provide points for writing output data.    At the first target load, the response is essentially elastic;  the second target load is sufficient to induce plastic strains on the order of 3\%. The target load information is given in the *.loads file.  For triaxial loading, three normal forces are specified for each target load consistent with the 
	desired stress state specified in Equation~\ref{eq:triaxial_stress_state}.}
\end{enumerate}
\subsubsection{Post-Processing:}
\begin{enumerate}
\item{{\bf Lightup:} Fiber-based quantities are computed for 6 fibers:  (100) and (111) crystal planes in the $[100], [010] {\rm and} [001]$ sample directions.}
\end{enumerate}
\subsubsection{Specifying options:}
\begin{enumerate}
\item{{\bf Load controls:}   The ``Triax-CSR'' mode is used to control the loading history.   Using this option, the magnitude of the velocities are adjusted to impose a specified loading rate (increase in the applied forces).  The relative values of the imposed surface velocities are adjusted to maintain the specified state of triaxial stress.    }
\item{{\bf Convergence criteria:} The default parameters have been used for both the velocity field and crystal stress iterative procedures.   The Newton-Raphson procedure is invoked for the velocity solutions.}
\item{{\bf Lightup:} The option to compute fiber-based quantities is specified.}
\end{enumerate}

\subsubsection{Selected Simulation Results}

\begin{figure}[h!]
\centering
\subfigure[ xx component]
{
\includegraphics*[width=6cm]{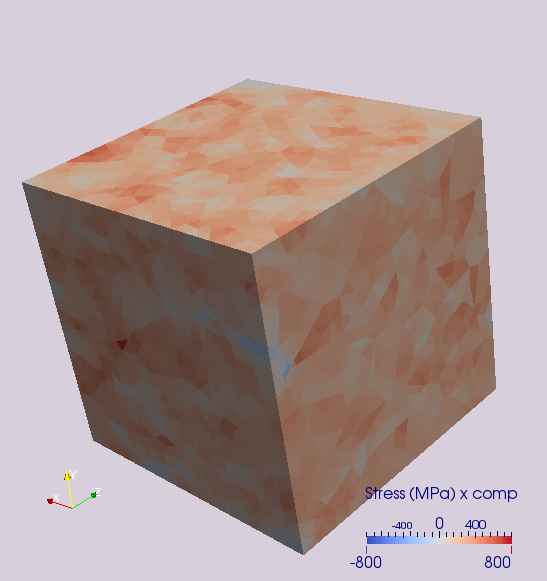} 
}
\subfigure[yy component]
{
\includegraphics*[width=6cm]{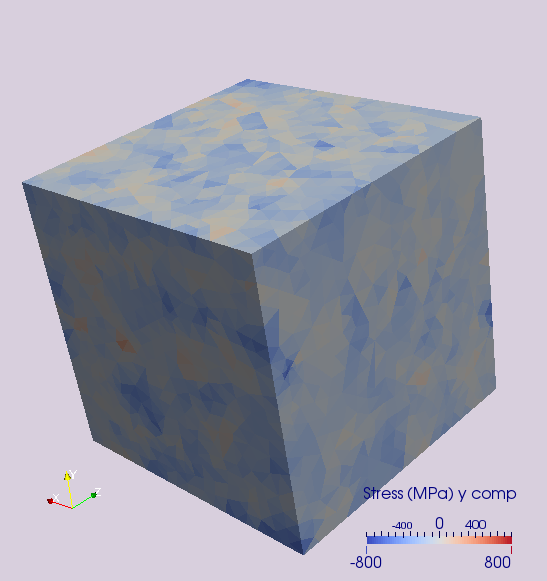}
}
\subfigure[zz component]
{
\includegraphics*[width=6cm]{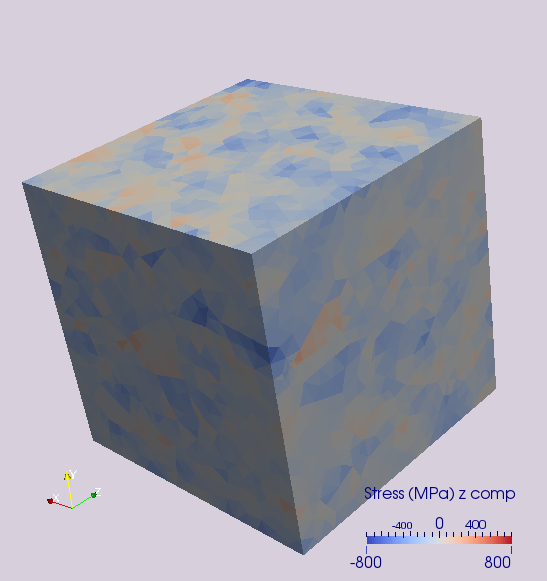}
}
\caption{Normal stress component distributions at $\sigma_{11} = 225$MPa}
\label{fig:example2_stresscomps}
\end{figure}

Stress distributions for the polycrystal are shown in  Figure~\ref{fig:example2_stresscomps} at the second target load. In these images, the normal components of the stress are   plotted over the deformed mesh.   The total deformation is not large, so the change in shape from the initial configuration shown in Figure~\ref{fig:example2_mesh_grains} is difficult to discern.  The differences in overall shade between the three images
reflects the triaxial stress condition intentionally imposed on the  polycrystal.   There are variations over the polycrystal for all of the components
stemming from the elastic and plastic anisotropy.  

\begin{figure}[h!]
\centering
\subfigure[ xx component]
{
\includegraphics*[width=6cm]{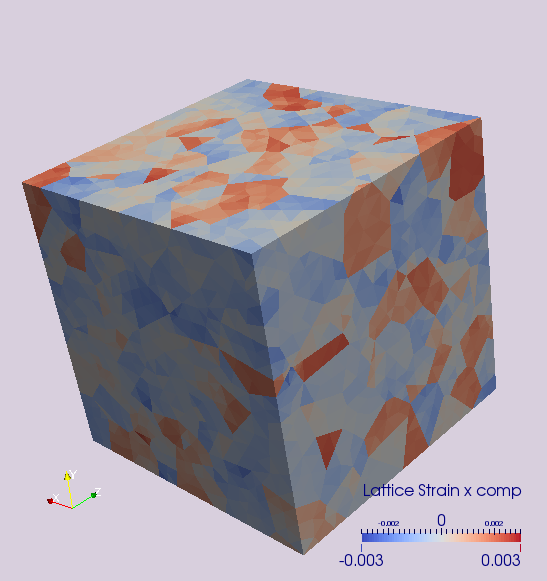} 
}
\subfigure[yy component]
{
\includegraphics*[width=6cm]{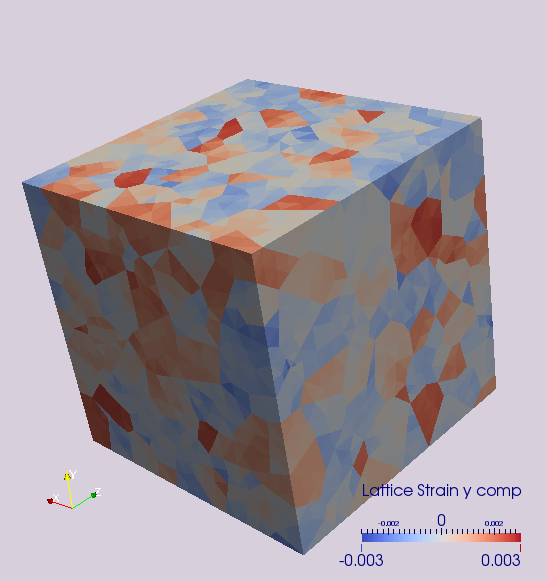}
}
\subfigure[zz component]
{
\includegraphics*[width=6cm]{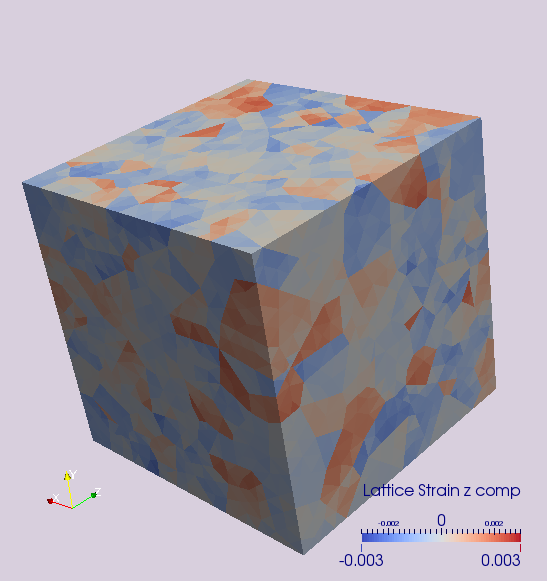}
}
\caption{Normal lattice (elastic) strain component distributions at $\sigma_{1} = 225$MPa}
\label{fig:example2_straincomps}
\end{figure}

\begin{figure}[h!]
\centering

\includegraphics*[width=10cm]{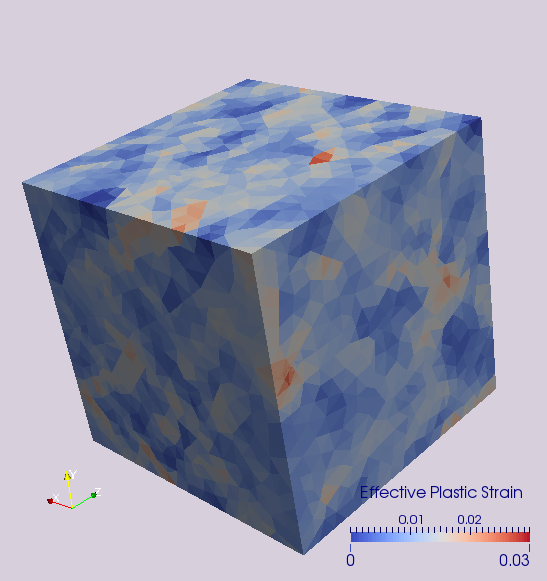} 

\caption{Effective plastic strain distribution at $\sigma_{1} = 225$MPa}
\label{fig:example2_plasticstrain}
\end{figure}

In Figure~\ref{fig:example2_straincomps} the normal components of the elastic strain are depicted.   A noticeable contrast to the stress distributions
is evident.   For the elastic strains, the net effect of the stress triaxiality and grain interactions is to produce distributions that span approximately the
same range in strain for all the normal strain components.  Unlike the stress, it is difficult to discern the triaxiality of the stress from differences in the lattice lattice (elastic) strain distributions.  
Figure~\ref{fig:example2_plasticstrain} shows the effective plastic strain at the second target load.  Again, the distribution is not uniform:  some elements display several percentage plastic strain while other have almost no plastic strain.
Details are available in \cite{mitch_ms}.

The distribution of stresses over a polycrystal depends strongly on the levels of anisotropy in the elastic and plastic behaviors of the constituent crystals.   The stress state must vary spatially to satisfy compatibility and equilibrium
if the properties vary.    Often these level are different, which is evident by observing the stress distributions as the polycrystal is loaded through the elastic-plastic transition.  At lower loads, the behavior is essentially purely elastic and the distribution is controlled by the elastic moduli.  At high loads, the behavior is dominated by the single crystal yield surface. This is illustrated by the relative changes in average elastic strains along the selected fibers given
in Table~\ref{tab:example2_fiberstrains}.  One can readily observe that between the target loads for $\sigma_{1} = 200$MPa and $\sigma_{1} = 225$MPa
there is not the same proportional increase in strains, as would be expected in the response remained linear and superposition could be applied. 
The adjustment in stress that accompanies yielding implies that the principal axes of the strain rotate as the stress moves toward a vertex of the single crystal yield surface.
 \begin{table}[h]
\centering
\begin{tabular}{||c|c|c|c|c|c|c||}
\hline
\hline
$\sigma_{1}$ (MPa) & $(100)||[100]$ & $(111)||[100]$  &  $(100)||[010]$ & $(111)||[010]$ & $(100)||[001]$ & $(111)||[001]$ \\
\hline
200 &  0.00211  & 0.00196 & -0.00115 &  -0.00117 &-0.00075 & -0.00078\\
 225 &  0.00221  & 0.00244 & -0.00105 & -0.00160 & -0.00044 & -0.00113 \\
\hline
\hline
\end{tabular}
\caption{Average lattice strains along selected fibers.}
\label{tab:example2_fiberstrains}
\end{table}

\pagebreak
   
\subsection{Voxel-Based Virtual Polycrystal Generated from 3-D Serial Section Maps}
\label{sec:voxelmesh}

This example was provided by Donald Boyce and was developed as part of an ONR-sponsored project on strength and ductility in titanium alloys. The example demonstrates the application of \fepx\, to a virtual polycrystal  that was built by mapping
voxel data to a regular mesh to define the grains. The titanium alloy being modeled is Ti-6Al-4V, which is two-phase at
room temperature.   However, since the volume fraction of the BCC phase is only about 7\%, this analysis is performed assuming a single (HCP) phase.

\subsubsection{Defining the virtual polycrystal}
\begin{enumerate}
\item{Phase Attributes:
The  principal phase for this titanium alloy has hexagonal symmetry (HCP), and is designated as the $\alpha$ phase.  The single crystal, hexagonal, elastic moduli for it are listed in Table~\ref{tab:example3_elasticmoduli}. 
Input to the code does not include $C_{33}$.  Rather, it is computed internally to assure that the constraint to decouple the
deviatoric and volumetric responses is satisfied.     The plasticity parameters were estimated from fitting stress-strain data for the alloy and are listed in Table~\ref{tab:example3_plasticparms}.   This type of alloy exhibits very little strain hardening; the choice of parameters accomplishes this by setting the initial slip system strength to the saturation value.   The slip systems include prismatic, basal and pyramidal systems as per Table~\ref{tab:slip_systems}.
This information is provided to \fepx\, in the *.matl input file.
 \begin{table}[h]
\centering
\begin{tabular}{||c|c|c|c|c|c||}
\hline
\hline
Phase &  Type & $C_{11}$ (GPa) & $C_{12}$ (GPa)  & $C_{13}$ (GPa)  &  $C_{44}$ (GPa) \\
\hline
$\alpha$ & HCP              & 161.4 & 91.0 & 69.5 & 46.7 \\
\hline
\hline
\end{tabular}
\caption{Titanium elastic moduli used in the single crystal constitutive equations for the voxel-based virtual polycrystal.}
\label{tab:example3_elasticmoduli}
\end{table}  
 \begin{table}[h]
\centering
\begin{tabular}{||c|c|c|c|c|c|c|c|c||}
\hline
\hline
Phase & $\gdotz$ (${{\rm s}^{-1}}$) & $m$ & $h_{0}$ (MPa) & $g_0$  (MPa) & $n^\prime$ & $g_1$ (MPa)   & $\gdots$ (${{\rm s}^{-1}}$) & $m^\prime$\\
\hline
$\alpha$ & 1.0 & 0.01 & 1000. &500. & 1  & 500. &  $5.0\times10^{10}$ & $0.01$  \\
\hline
\hline
\end{tabular}
\caption{Titanium slip parameters used in the single crystal constitutive equations for the voxel-based virtual polycrystal.}
\label{tab:example3_plasticparms}
\end{table}
 \begin{table}[h]
\centering
\begin{tabular}{||c|c|c||}
\hline
\hline
Basal & Prismatic & Pyramidal\\
\hline
  1 & 1 & 3  \\
\hline
\hline
\end{tabular}
\caption{Relative strength for the titanium slip system used in the single crystal constitutive equations for the voxel-based virtual polycrystal.}
\label{tab:example3_ss-strengths}
\end{table}
}
\item{{\bf Mesh definition:} 
The finite element mesh underlying the virtual polycrystal was instantiated by mapping voxel-based (3D) orientation map onto a regular mesh using a custom \matlab\, script (available in the \odfpf\, package).   The orientation map was obtained from serial section data measured using electron back-scattered diffraction (EBSD).     The finite element mesh spans a volume of $20\mu m \times 20 \mu m \times 60 \mu m$ that 
coincides with an interior portion of the experimental volume.  The element size was chosen to give and resolution comparable to the spatial resolution of the data.   
The resulting mesh, shown in Figure~\ref{fig:example3_mesh}, has 144,000 10-node tetrahedral elements and 230,401 nodal points.
The corresponding arrays for the nodal point coordinates and element connectivities are  provided to \fepx\, in the *.mesh input file, along with 
definition of the six sample surfaces in terms of the mesh elements.  
 \begin{figure}[h!]
\centering
\subfigure[Finite element mesh]
{
\includegraphics*[width=6cm]{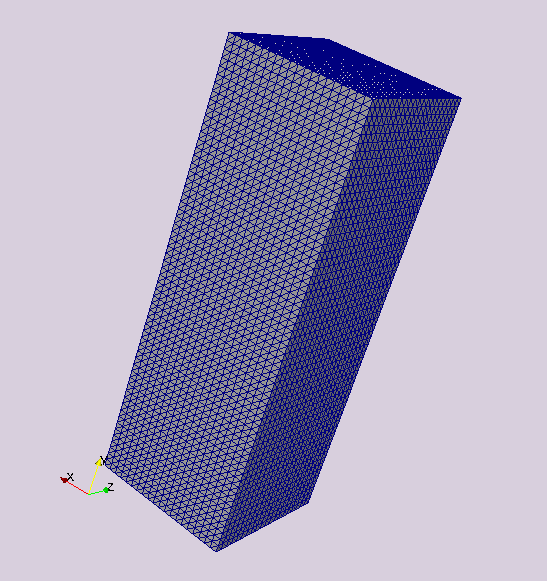} 
}
\subfigure[Lattice orientations]
{
 \includegraphics*[width=6cm]{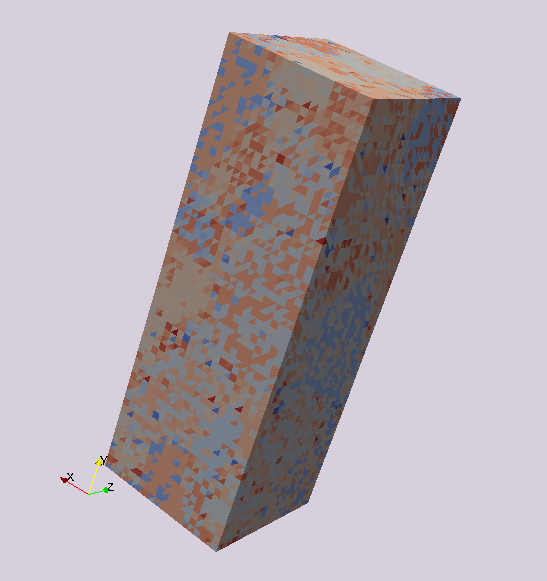} 
 }
\caption{Finite element mesh for the mill annealed titanium alloy and grain orientations assigned to the elements.}
\label{fig:example3_mesh}
\end{figure}
}

\item{{\bf Phase and grain definition:}
For every element of the mesh, the voxel that is closest to the element is identified using the distance between the centroid of the element and the centroid of the voxel). The orientation data of the voxel is then used to assign the lattice orientation for the element.  
 The simulation assumes all grains are of the same HCP phase.  
All the grains are assigned the same initial slip system strength, which in turn was assigned to every element of the grain.  
Figure~\ref{fig:example3_mesh} also shows the grain assignment associated with the mesh.
While some grains are evident, noisy or missing information in the orientation data makes crisp definition of grains difficult.
The grain assignment and lattice orientation data are given in the *.grain and *.kocks files. 
}
\item{{\bf Vertex files:} 
Vertices of a HCP single crystal yield surface was used having a topology consistent with 
the prescribed ratios of the slip system strengths of $(1:1:3)$ for the basal, prismatic and pyramidal slip systems, respectively.
} 
\end{enumerate}

\subsubsection{Controlling the loading:}
\begin{enumerate}
\item{{\bf Boundary conditions:} 
Boundary conditions are chosen to mimic a  tensile test:   there is a fixed velocity applied on the upper  surface (positive $y$ face) while the lower surface is fixed from translation in the $y$ direction.  The lateral surfaces have 
two traction free surfaces (positive $x$ and $z$) and two symmetry surfaces (negative $x$ and $z$).
This information is in the *.bcs file.
}
\item{{\bf Target loads:} 
Simple load control is applied to compress the sample.   Three y-direction target loads along a monotonically increasing path (no unloading episodes) are specified to provide points for writing output data. The final target load was sufficient to compress the sample by approximately 1\%, overall.   This information is given in the  *.loads file.
}
\end{enumerate}

\subsubsection{Specifying options (information in the *.options file):}
\begin{enumerate}
\item{{\bf Load controls:  }  The ``control by load'' mode is used to control the loading history.
}
\item{{\bf Convergence criteria: }   Default parameters have been used for both the velocity field and crystal stress iterative procedures.  The Newton-Raphson procedure is invoked for the velocity solutions.
}
\end{enumerate}

\subsubsection{Selected Simulation Results}  
   
\begin{figure}[h!]
\centering
\subfigure[Axial stress ]
{
\includegraphics*[width=6cm]{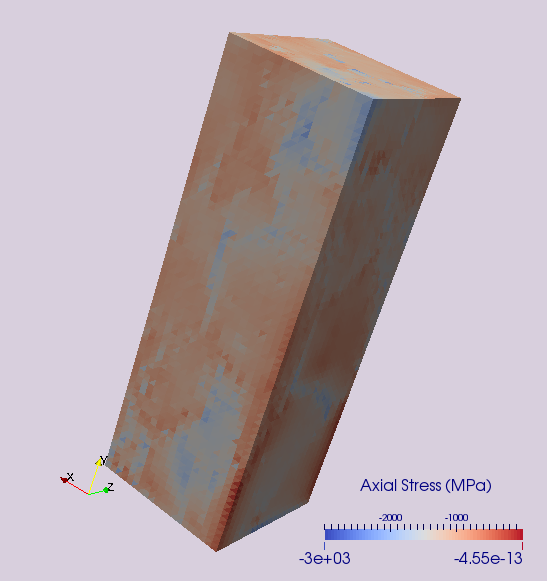} 
}
\subfigure[Axial lattice (elastic) strain]
{
\includegraphics*[width=6cm]{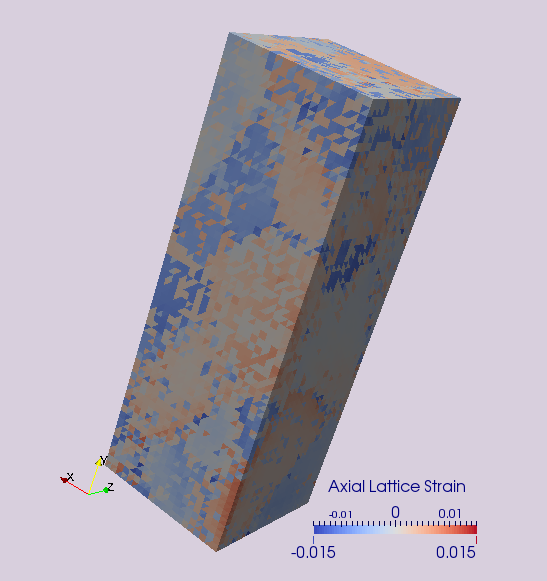} 
}
\caption{Axial stress and axial lattice strain distributions at the third target load.}
\label{fig:example3_axialstress}
\end{figure}

Figure~\ref{fig:example3_axialstress} shows the axial stress  and the axial lattice strain at the final target load.  
Here the grain structure is more evident.  
Because the load is sufficient to cause wide-spread yielding, higher stresses and strains typically occur in grains whose lattices are at stronger orientations.
Note that the stress distribution is not merely a scaled version of the strain distribution.  This is a result of the elastic anisotropy, which implies that the eigenvectors of the stress and strain tensors do not necessarily align.

\begin{figure}[h!]
\centering

\subfigure[Mean stress]
{
\includegraphics*[width=6cm]{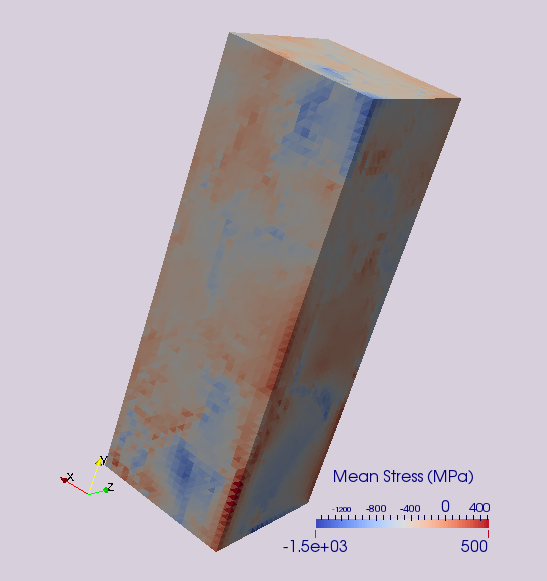}
}
\subfigure[Effective plastic strain]
{
\includegraphics*[width=6cm]{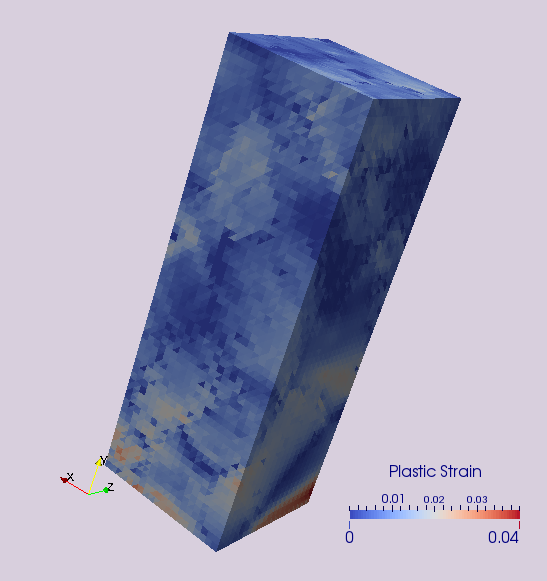}
}
\caption{Mean stress  and plastic strain distributions at the third target load.}
\label{fig:example3_plasticstrainrate}
\end{figure}

At first glance, the stress levels depicted in Figure~\ref{fig:example3_axialstress} might appear to exceed stress limits imposed the single crystal yield surface.   However, the large grain size relative to the sample size has an effect such that the deformation is more highly constrained.
The consequence is that the mean stress (shown in Figure~\ref{fig:example3_plasticstrainrate})  is larger at many points than would be expected for a case of simple tension.  
The plastic straining induced by this highly heterogeneous stress field is very inhomogeneous, as is evident from the distribution of the effective plastic strain  shown in  
Figure~\ref{fig:example3_plasticstrainrate}.  This plot shows how plastic flow  interconnects through the polycrystal leaving some domains relatively undeformed.  

\pagebreak[4]
\section{Acknowledgments}
Support was provided  by the US Office of Naval Research (ONR) under contract N00014-12-1-0075.   The authors wish to thank Andrew Poshadel and Amand Mitch for the example problems they provided.  Thanks also to Andrew Poshadel, Matthew Kasemer and Robert Carson for their comments on the manuscript.
\newpage
\bibliographystyle{unsrt}
\bibliography{FEpX.bib}
%
%  ==================== End Document
%
\end{document}